\documentclass[aps,twocolumn,floats,prl,footinbib]{revtex4-1}
\usepackage{graphics,graphicx,epsfig}
\usepackage{amssymb,color}
\usepackage{epsf,epstopdf,wrapfig}
\usepackage{amsmath}
\usepackage{amsmath}
\usepackage{amssymb}
\usepackage{graphicx}
\usepackage{bm}
\usepackage{xr}
\usepackage{scrextend}

\usepackage{lipsum}  

\usepackage{amsfonts}
\usepackage{color}
\newcommand{\de}{\partial}
\newcommand{\vv}{\boldsymbol{v}}
\newcommand{\vk}{\boldsymbol{k}}
\newcommand{\vx}{\boldsymbol{x}}

\begin{document}
	
	\title{Evidence of fluctuation-induced first-order phase transition in active matter}
	\author{
		Luca Di Carlo$^{2,1}$
	}
	\author{
		Mattia Scandolo$^{2,1}$
	}
	\affiliation{$^2$ Istituto Sistemi Complessi, Consiglio Nazionale delle Ricerche, UOS Sapienza, 00185 Rome, Italy}
	\affiliation{$^1$ Dipartimento di Fisica, Universit\`a\ la Sapienza, 00185 Rome, Italy}
	\begin{abstract}
		We study the properties of the Malthusian Toner-Tu theory in its near ordering phase. Because of the birth/death process, characteristic of this Malthusian model, density fluctuations are partially suppressed. 
		We study this model using the perturbative renormalization group. At one loop we find that the renormalization group flow drives the system in an unstable region, suggesting a fluctuation-induced first-order phase transition.
	\end{abstract}
	
	\maketitle
Despite its simple definition, the Vicsek model \cite{vicsek+al_95,vicsek_review} represents a challenging problem.
Its rich phenomenology and complexity arise from the combined effect of the particles' ferromagnetic/alignment interaction and activity, that is particles' self-propulsion.
The consequent rewiring of the interaction network implies feedback between density and velocity fluctuations, which turns an equilibrium second-order phase transition into an off-equilibrium first-order phase transition \cite{baglietto2009nature,chate+al_08b}. 
The nature of this phase transition can be understood as a liquid-gas phase transition \cite{Solon2015} and both numerical simulations and theoretical studies show that density fluctuations play a crucial role in determining the properties of the transitions in active matter \cite{baglietto+al_12, ginelli2016physics,Martin2021FIPS,mishra2022active,cates2015motility}.

A possible way to tackle this problem, and study the transition in the Vicsek model,  consists of considering its corresponding hydrodynamic theory: the Toner-Tu (TT) theory \cite{toner1998flocks}.
However, precisely because of the velocity-density coupling, deriving the properties of the TT theory in its near-critical regime is a highly complicated task.
In the seminal paper \cite{chen2015critical}, Chen et al. studied the critical dynamics of the TT theory in the incompressible limit (ITT) using the perturbative renormalization group (RG) - one of the first RG calculations showing that active matter systems can belong to a  universality class of their own. They found that, if density fluctuations are \textit{completely} suppressed, the phase transition is second-order, and a novel active off-equilibrium fixed-point rules the critical dynamics \cite{chen2015critical,cavagna21equilibrium}.
If, on the one hand, this calculation shows the relevance of activity in the critical dynamics, on the other hand, it ignores \textit{completely} density-velocity coupling; for this reason, it cannot explain why, and how, the order/disorder phase transition becomes first order.

A renormalization group calculation that explains the role of density fluctuations in turning the phase transition from second to first order is still missing. Here, we  try to address this problem by studying the critical dynamics of a modification of the TT theory, namely the Malthusian Toner-Tu (MTT) theory \cite{toner2012birth}. 
In the MTT theory, density fluctuations are \textit{ partially} suppressed, and in this sense, MTT theory poses in between the ITT and the TT theories. This theory is easy enough to be studied with perturbative renormalization group techniques, but it is still rich enough to produce a behavior qualitatively different from the one of the incompressible theory. 

The MTT theory, introduced by J. Toner in \cite{toner2012birth}, is a model for self-propelled agents that reproduce and die. In its hydrodynamic description the equations of motion (EOM) for the velocity field $\vv$ and the density field $\rho$, as introduced in \cite{toner2012birth}, are:

\begin{align}
	\de_t \vv+ \gamma_{1} \vv\cdot \nabla \vv  &=-\frac {d V[\vv]} {d \vv} + \Gamma_0 \nabla^2 \vv  -\nabla P + \boldsymbol{f} +\dots \label{babbala}\\
	\de_t \rho + \nabla\cdot(\rho \vv) &= g(\rho) \label{ariba} \quad  \ 
	V  [\vv] = \int d^d x \ \frac 12 m_0 \vv^2  + \frac 14 J_0 \vv^4 
\end{align}
where $P$ is a pressure term, $\boldsymbol{f}$ is a Gaussian white noise with variance $2 D_0$, and $(\vv\cdot\nabla)\vv$ is the standard material derivative \cite{landau_fluids}.
The term $g(\rho)$ accounts for the birth/death process and it tends to keep density near to a fixed value $\rho_0$ \cite{toner2012birth,chen2020universality}. Near to $\rho_0$ the function $g(\rho)$ acts as a harmonic restoring force, $\de_t \rho = g^\prime(\rho_0) \delta \rho$, where $\delta \rho = \rho-\rho_0$, $g(\rho_0 ) = 0 $, and $g^\prime( \rho_0)  <0$.
At first glance the birth/death mechanism seems to further complicate the problem, but in fact it is a significant simplification. Because of the term $g(\rho)$, density is not a conserved field \cite{toner2012birth}, $\de_t \rho(\vk= 0,t)\neq 0$, and in this sense $\rho$ is not an hydrodynamic variable anymore \cite{ramaswamy_review}. Following \cite{toner2012birth,chen2020moving,chen2020movingprl}, from the continuity equation \eqref{ariba} it is possible to derive a \textit{linear} relation between density and velocity fluctuations:
\begin{equation} 
	\delta \rho(\vx,t) =  -\frac{\rho_0}{|g^\prime(\rho_0)|} \ 	\nabla \cdot \vv(\vx,t) 
	\label{ingannodellacadrega}
\end{equation} 

The force term $\boldsymbol{F}(\vv) = -d V/d \vv$ is responsible for the order/disorder phase transition, and it arises from the imitation/ferromagnetic interaction. The state of the system depends on the values of $m_0$ and $J_0$ \cite{toner_review}. While the coupling $J_0$ must be positive - otherwise the potential $V$ would be unbounded - 
the sign of $m_0$ determines, at the mean-field level, whether or not the system is in a polarized phase $\langle \vv \rangle \neq 0$. Precisely, if $m_0<0$ the average polarization of the system is non-zero,  $|\langle  \vv \rangle| = \sqrt{-m_0 /J_0}$.

Before proceeding with the explicit derivation of the EOM, a remark is in order. In general, the TT theory breaks Galilean invariance \cite{landau_fluids,toner_12}, and the Malthusian TT theory is no exception; for this reason, we must include in the model all the possible terms with one derivative and two fields, namely $\vv \nabla \cdot \vv$  and $\nabla \vv^2$  \cite{toner_12}.
The first of these two terms can be interpreted as a density-dependent alignment force \cite{Martin2021FIPS}; assuming that the alignment force $m_0 \vv $ depends on density, and using the relation \eqref{ingannodellacadrega}, leads to:
\begin{equation}
	m(\rho) \vv = m(\rho_0) \vv -m^\prime(\rho_0) \frac{\rho_0}{|g^\prime(\rho_0)|} \vv  \nabla \cdot \vv
\end{equation}
where $m(\rho_0)$ is simply $m_0$, and the second term reproduces the expected structure $\vv \nabla\cdot \vv$. The second non-Galilean-invariant term can be interpreted as a non-linear pressure term of the kind $P \sim - \gamma_3 \vv^2 $. This would lead to the following pressure force, 
\begin{equation}
	-\nabla P(\rho,\vv) = -\nabla P (\rho) + \gamma_3 \nabla \vv^2 \ \ , 
\end{equation}
which reproduces the structure of the second non-Galilean-invariant term, $\nabla \vv^2$.

To derive the EOM of the model we need an explicit expression for the pressure force $\nabla P$. 
Following the literature on active matter theories \cite{toner_review}, the pressure term $P$ can be expressed as a series in powers of density fluctuations:
\begin{equation}
	P(\rho) = \sum_{n = 1} \sigma_n \delta \rho ^n \ .
	\label{underpressure} 
\end{equation}
Using equation \eqref{ingannodellacadrega}, the pressure force, at the leading order in the fluctuations, reduces to
\begin{equation}
	P\simeq \sigma_1 \delta \rho = -\sigma_1 \frac{\rho_0}{|g^\prime (\rho_0)|} \nabla \cdot \vv(\vx,t) \ \ .
	\label{ilpotentissimo}
\end{equation}
This leads to the following EOM for the velocity field, 
\begin{equation}
	\begin{split}
		&\de_ t \vv + \gamma_1 (\vv \cdot \nabla) \vv = -(m_0 + J_0 \vv^2) \vv + \Gamma_0 \nabla^2 \vv  \\& +\gamma_2 (\nabla \cdot \vv) \vv + \gamma_3 \nabla v^2+\sigma_1\frac{\rho_0}{|g^\prime ( \rho_0)|} \nabla( \nabla \cdot \vv) + \boldsymbol f \ \ ,
	\end{split} 
	\label{adelante}
\end{equation}
\begin{equation} 
	\langle f_{\alpha} (\vx,t) f_\beta(\vx^\prime,t^\prime )\rangle = 2 D_0 \delta^d(\vx-\vx^\prime) \delta(t-t^\prime) \ \ .
\end{equation}

We remark that we obtained an EOM for the sole velocity field; however, a relic of the coupling with density is hidden in the parameter $\gamma_2$, whose interpretation is that of a density-dependent ferromagnetic interaction. 
Equation \eqref{adelante} is the starting point of the perturbative renormalization group calculation.

Before starting to study the properties of the Malthusian theory near the transition, it is helpful to study its linear limit. 
The linearized EOM, 
in Fourier ($\vk,t$) space, is: 
\begin{equation}
	\begin{split}
		\de_ t v_\alpha(\vk,t) = -m_0 v_\alpha(\vk,t)- \Gamma_0 \vk^2 v_\alpha(\vk,t)\\ 
		-\sigma_1 \frac{\rho_0}{|g^\prime(\rho_0)|} k_\alpha k_\beta v_\beta(\vk,t) + f_\alpha(\vk,t) 
	\end{split}
	\label{balilla} 
\end{equation}
It is evident that, even at the liner level, the EOM \eqref{balilla} is anisotropic. A part of the force is in the direction of the field $v_{\alpha}$, while another part of the force is in the direction of the wave-vector $k_\alpha$.  
For this reason, we have to distinguish between longitudinal and transverse fluctuations, namely fluctuations parallel and orthogonal to the $\hat {\boldsymbol{ k}}$ direction. We define the transverse $P^{\perp}_{\alpha \beta} (k)$ and longitudinal $P^\parallel _{\alpha \beta} (k)$ projection operators: 
\begin{equation}	P^\perp_{\alpha \beta}(\boldsymbol k) = \delta_{\alpha \beta} - {k_\alpha k _\beta} /{k^2} \qquad , \quad P^ {\parallel}_{\alpha \beta} (\boldsymbol k) = {k_\alpha k_\beta} /{k^2} \  \mbox{,}
\end{equation} 
We define also the longitudinal $\Gamma_0^\parallel$ and transverse  $\Gamma_0 ^\perp$ kinetic coefficients,
\begin{equation}
	\Gamma_0 ^\perp = \Gamma_0 \qquad \quad\quad  \Gamma_0 ^\parallel = \Gamma_0 + \sigma_1
	\frac{ \rho_0}{|g^\prime(\rho_0)|} \ \mbox{,}
	\label{dwight}
\end{equation}
which in general are not equal. The responsible for their difference is the pressure force, indeed $(\Gamma_0 ^\parallel - \Gamma_0 ^\perp)$ is proportional to $\sigma_1$. Intuitively, since longitudinal fluctuations are coupled to density fluctuations, as shown in equation \eqref{ingannodellacadrega}, they relax faster than transverse fluctuations and $\Gamma_0^\parallel > \Gamma_0  ^\perp$. From the equation of motion \eqref{balilla} it is possible to derive the linear response and correlation functions $G_{\alpha\beta} ^0(\vk,\omega)$ and $C_{\alpha\beta} ^0 (\vk,\omega)$.
\begin{align}
	G^0_{\alpha \beta}(\vk,\omega)  &= G_0 ^\perp(\vk,\omega) P^\perp _{\alpha \beta}(\vk) + G^{\parallel}_0 (\vk,\omega) P^\parallel _{\alpha \beta }(\vk)   
	\label{freezer}\\
	C^0_{\alpha \beta} (\vk,\omega)  &= C_0^\perp(\vk,\omega) P^\perp _{\alpha \beta}(\vk) + C_0 ^ {\parallel}(\vk,\omega) P^\parallel _{\alpha \beta } (\vk)
	\label{cell}
\end{align}
where the functions $G_0 ^{\parallel,\perp}(\vk,\omega)$ and $C_0 ^{\parallel,\perp} (\vk,\omega)$ are defined as follows
\begin{align}
	G^{\perp,\parallel} _0(\vk,\omega)&= \frac{1}{-i\omega + \Gamma^{\perp,\parallel} _0 \vk^2+m_0}
	\label{kaioken}
	\\
	C^{\perp,\parallel}_0 (\vk,\omega) &= \frac{2 D_0}{\omega^2 + (\Gamma^{\perp,\parallel}_0 \vk^2+m_0)^2}  \ \mbox{.}
	\label{kaiokenx20}
\end{align}
These expressions, \eqref{kaioken} and \eqref{kaiokenx20}, can be obtained with standard procedures \cite{cardy1996scaling}, and a detailed derivation can be found in the SM. 

The ITT theory \cite{chen2015critical} can be formally recovered in the $\Gamma^\parallel _0 \gg \Gamma^\perp _0 $ limit; in this regime, both $G_0 ^\parallel$ and $C_0 ^\parallel$ are negligible, and the bare correlation and response functions are proportional to $P_{\alpha \beta} ^\perp (\vk)$, as in incompressible theories \cite{chen2015critical,foster77}.

\textit{Renromalization group equations - } The momentum-shell RG approach \cite{cardy1996scaling} unfolds through two stages: \textit{i)} integrating out small length-scale details, on the so-called momentum shell $\Lambda/b < k <\Lambda$, where $b \simeq 1$; \textit{ii)} rescaling of momenta, $ \vk^\prime = \vk/ b$, frequencies,  $\omega^\prime = \omega/b^z$, and fields, $\vv^\prime (\vk,\omega) = b^{d_{ v}}  \vv(  b^{-1}\vk,b^{-z}\omega)$. The effect of step \textit{ii} is to rescale each parameter by its corresponding naive dimension.
The effect of step \textit{i} is to provide each parameter with a perturbative scaling dimension, which is computed using the Feynman diagrams technique (the technical details can be found in the SM). 
Combining the perturbative and the naive scaling dimensions leads to the following set of recursion relations, which determine how the parameters of the model change when we observe  the system at larger and larger length scales \cite{cardy1996scaling}:
\begin{align}
	& m_b = m_0 \ b^{z}( 1 + \delta m \ln b)  \label{patato}\\
	& \Gamma^{\parallel, \perp}_b = \Gamma_0^{\parallel, \perp}\  b^{z-2 -\eta} (1+\delta \Gamma^{\parallel, \perp}\ln b ) \\
	&\gamma_{i,b} = \gamma_{i,0} \ b^{d_v-d-1-\eta} (1 + \delta\gamma_i \ln b) \\
	&  J_b = J_0 \ b^{2d_v -z-2d-\eta}(1+\delta J \ln b) \\ 
	& D_b = D_0 \ b^{-2 d_{ v} +d+z-2 \eta }(1+ \delta D \ln b) \label{patoto}
\end{align}
Here,  $\delta X$ represents the perturbative correction to the parameter $X$, and $\eta$ is the perturbative anomalous dimension of the field. A set of 15 Feynman diagrams determines the perturbative contributions - their explicit expressions can be found in the SM. Iterating the RG transformation results in a flow in the space of theories: the \textit{renormalization group flow}.

It is particularly useful to define a set of effective parameters and coupling constants,
\begin{equation}
	\begin{split}
		g_{1,0} & = \frac{\gamma_{1,0}}{\Gamma_0^\perp} \sqrt{ \frac{D_0}{\Gamma_0^\perp }}\  \quad g_{2,0} = \frac{\gamma_{2,0}}{\Gamma_0^\perp} \sqrt{ \frac{D_0}{\Gamma_0 ^\perp}} \ \quad g_{3,0}= \frac{\gamma_{3,0}}{\Gamma_0^\perp} \sqrt{ \frac{D_0}{\Gamma_0 ^ \perp}}\\ 
		u_0 &= \frac{ J_0 }{\Gamma_0 ^\perp } \frac{ D_0}{\Gamma_0 ^\perp  }  \ \ \qquad \qquad \mu_0 =   \frac{\Gamma_0 ^\parallel}{\Gamma_0 ^ \perp}   \qquad  \ \ \ \ r_0 = \frac{m_0}{\Gamma^\perp _0} 
	\end{split}
	\label{effective_par_def}
\end{equation}
where $r_0$ and $u_0$ are the effective couplings relative to the force term $m_0\vv+J_0 \vv^2 \vv$, and $g_{i,0}$ are the effective couplings relative to the three non-linearities, $\vv\cdot \nabla \vv$, $\vv \nabla \cdot \vv$ and $\nabla \vv^2$. The parameter $\mu$ is the ratio between the longitudinal and transverse kinetic coefficients, and it determines how the longitudinal fluctuations relax faster than the transverse fluctuations. 
The renormalization group flow of these parameters determines the large scale properties of the system. 
The recursion relations for the effective parameters can be derived by the relations \eqref{patato}-\eqref{patoto}:

\begin{align}
	& r _b = r_0 b^2 \left[ 1 + (\delta m_0-\delta \Gamma ^\perp) \ln b \right] \\
	&\mu_b = \mu_0 \left[1+ (\delta \Gamma^\parallel  - \delta \Gamma^\perp )\ln b\right] \label{mellow} \\
	&u_b =  u_0 b^{\epsilon} \left[ 1 + (\delta J + \delta D -2 \delta \Gamma^\perp -\eta) \ln b \right]\\
	&g_{i,b}   = g_{i,0} b^{\frac \epsilon 2} \left[1+ \left (\delta \gamma_i + \frac 12 \delta D- \frac 32 \delta \Gamma^\perp-\eta\right) \ln b\right] \label{mood}
\end{align}

\par  The naive scaling dimensions of the effective parameters are:
\begin{equation}
	d_u ^{(n)} = \epsilon \qquad d_{g} ^{(n)} = \epsilon /2 \qquad d_\mu ^{(n)} = 0 \ \mbox{,}
\end{equation}
where $\epsilon$ is the distance from the upper critical dimension $\epsilon = 4-d$. 
This means that in dimension $d<4$ both $u$ and the constants $g_i$ are relevant in the renormalization group sense.

\textit{Incompressible case - }
In the  $\Gamma^\parallel\gg \Gamma^\perp$ limit, the bare correlation functions of the Malthusian TT theory coincide with the ones of an incompressible theory \cite{foster77,chen2015critical}. 
Therefore, in this limit it is reasonable to expect that the transition is ruled by the exponents of incompressible active matter found in \cite{chen2015critical}.
For $\Gamma^\parallel\gg\Gamma^\perp$, which implies $\mu\to\infty$, the recursion relations \eqref{kaioken}-\eqref{kaiokenx20} reduce to the ones of \cite{chen2015critical} (see SM). The large scale behavior for $\mu \to \infty$ is thus ruled by the critical exponents,
\begin{equation}
	z =2 -\frac {31}{113} \epsilon \quad \qquad \beta = \frac 12 - \frac 6{113}\epsilon  \qquad \nu = \frac 12 +  \frac{29}{226} \epsilon  \ , 
	\label{spocrit} 
\end{equation}
which coincide with the ones of incompressible active matter \cite{chen2015critical}.
However, for large but finite values of $\mu$, namely in the neighborhood of the incompressible fixed point, the recursion relation of $\mu$ is 
\begin{equation}
	\mu_b = \mu_0\left[ 1 -\frac {g_{1}^2}{4} \ln b \right] = \mu_0 b^{-g_1^2/4} \quad \quad \mbox{as} \ \ \  \mu  \gg1  \ \mbox{,}
\end{equation}	
revealing that the incompressible fixed point is unstable under a small deviation from incompressibility. This means that, for large but finite $\mu$, the RG flow escapes this fixed point. Moreover, the larger $\mu$, the longer the RG flow lingers near the incompressible active matter fixed point; this structure gives rise to a crossover  \cite{cardy1996scaling}. To determine whether or not the critical dynamics is ruled by the incompressible fixed point we must consider the stopping condition of the RG flow. The RG flow stops when the correlation length, which trivially scales as $\xi_{b} = \xi/b$, is of the order of the  inverse cutoff $\Lambda^{-1}$, namely when $\xi_{\mathrm{stop}} = \Lambda^{-1}$, 
where $l_{\mathrm{stop}} = \log_b (\Lambda \xi) $ is the number of RG iterations before stopping. If at the end of the RG flow $\mu_{\mathrm{stop}}$ is still large, the incompressible active matter fixed point rules the critical dynamics. For small enough correlation length, $\xi \ll \mu_0^{\phi}$, 
the critical dynamics is ruled by the incompressible active matter fixed point \eqref{spocrit} - where the crossover exponent \cite{cardy1996scaling} is 
\begin{equation}
\phi =\frac{31}{113} \epsilon.
\end{equation}
In finite systems the correlation length is always bounded by the system size $L$.
Hence, if $L \ll \mu ^{\phi}$ the critical dynamics is ruled by the incompressible fixed point, meaning that the results of Chen \textit{et al.} \cite{chen2015critical} also hold in sufficiently small systems with mild density fluctuations $\mu\gg1$. Conversely, if the system size is large enough, $L \gg \mu ^{\phi}$, its properties at the transition point are not described by the incompressible active matter fixed point. In that case we must study the recursion relations for finite $\mu$. We remark that $\mu$ can be measured by looking at the system's transverse and longitudinal correlation functions, and, at least in principle, it could be measured both in numerical simulations and real experiments.

\begin{figure} [t]
	\centering
	\includegraphics[width = 0.48 \textwidth]{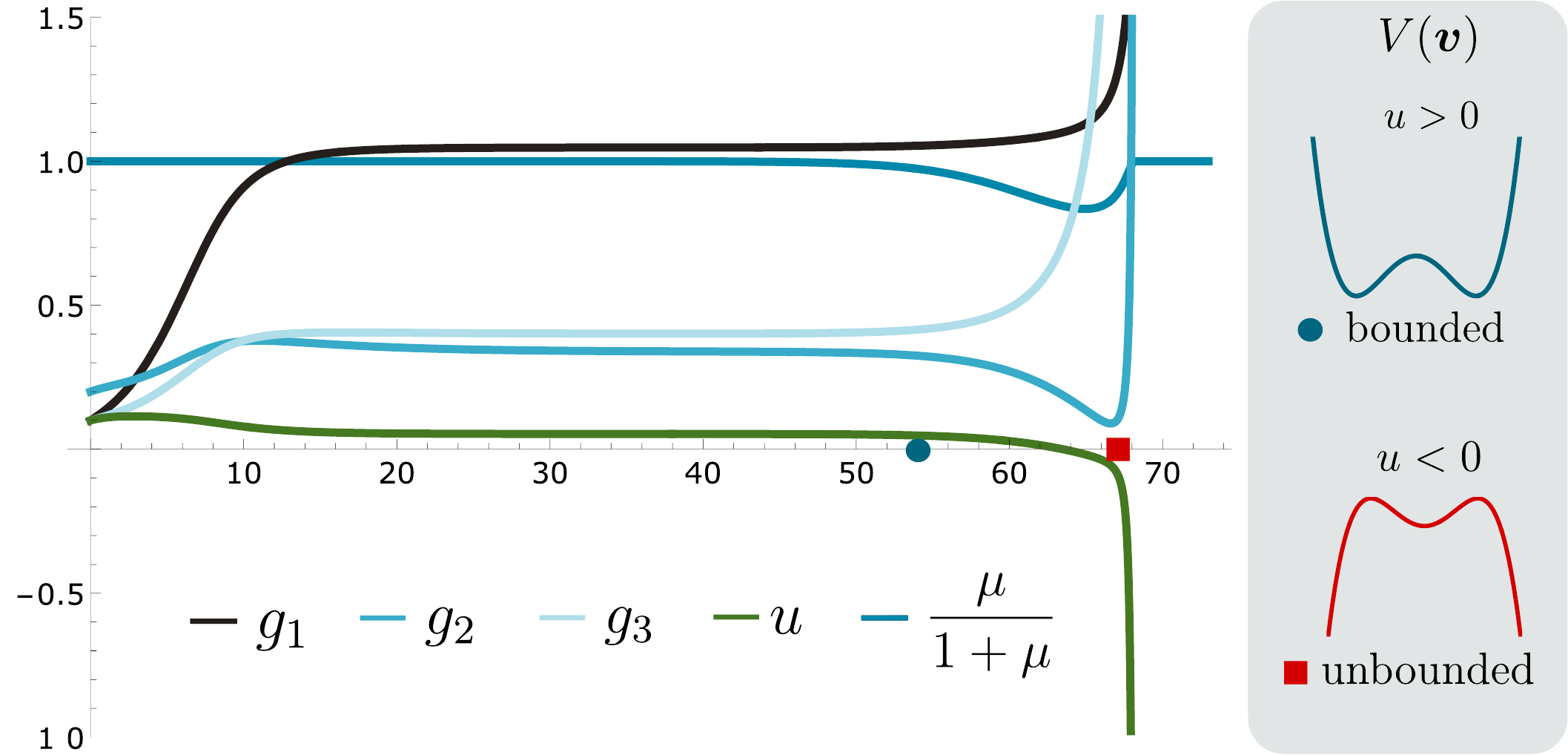}
	\caption{Renormalization group flow of the Malthusian Toner-Tu theory. In this example, the flow starts in the nearly incompressible regime, $\mu\gg1$ and $\mu/(1+\mu)\simeq 1 $; as soon as the flow escapes from the incompressible regime, the ferromagnetic coupling $u_b$ (green curve)  becomes negative, and then the flow is characterized by run-away trajectories. In the $u_b<0$ regime the system is unstable. In the inserts we show qualitatively how the pseudo-potential $V(\vv)$  changes before (blue dot) and after (red square) the flow enters the unstable region; when $u_b$ becomes negative the pseudo-potential is unbounded.
	}
	\label{gattiniEsplosivi}
\end{figure}

\textit{Fluctuation-induced first-order phase transition - }
The properties of the Malthusian TT theory in the vicinity of the transition are determined by the recursion relation \eqref{mellow}-\eqref{mood}, see SM.
These recursion relations are complicated, and it is not possible to study them analytically. For this reason, we choose a set of initial values of the parameters and simulate the recursion relations \eqref{mellow}-\eqref{mood} numerically. Usually, this procedure leads the system's parameters to an infrared-stable fixed point; however, things are more complicated in this model. In figure  \ref{gattiniEsplosivi} we show the renormalization group flow obtained by numerical integration; we chose a large initial value of $\mu$, meaning that the RG flow starts in a neighborhood of the incompressible active matter fixed point. As soon as the flow escapes 
this fixed point - which means that the longitudinal fluctuations, along with density fluctuations, become relevant - the RG flow enters an unstable region, where the ferromagnetic coupling $u_b$ becomes negative. Moreover, after that, the RG flow is characterized by run-away trajectories, and the couplings blow up to larger and larger values. The condition $u>0$ ensures that the pseudo-potential $V(\vv)$ is bounded, and as shown in figure \ref{gattiniEsplosivi} after the potential enters the $u_b<0$ region, the potential becomes unbounded. 

To have an idea of what is happening, we consider how the pseudo-potential $V(\vv)$ changes along the RG flow. At the beginning of the RG flow, the pseudo-potential $V(\vv)$ takes a double-well shape, characteristic of $\lambda \varphi^4$ theories \cite{cardy1996scaling}, guaranteeing that the order parameter is small. However, if the flow enters the $u_b<0$ region, the pseudo-potential becomes unbounded. In this regime, the order parameter is clearly not fluctuating around zero anymore, suggesting a breakdown of the perturbative expansion. The situation may seem puzzling: all the theories along an RG trajectory describe the same system, which means that if the system is stable at the beginning of the flow, it must remain stable along the whole RG trajectory, but figure \ref{gattiniEsplosivi} clearly shows the contrary.

\begin{figure}[t]
	\includegraphics[width = 0.45\textwidth]{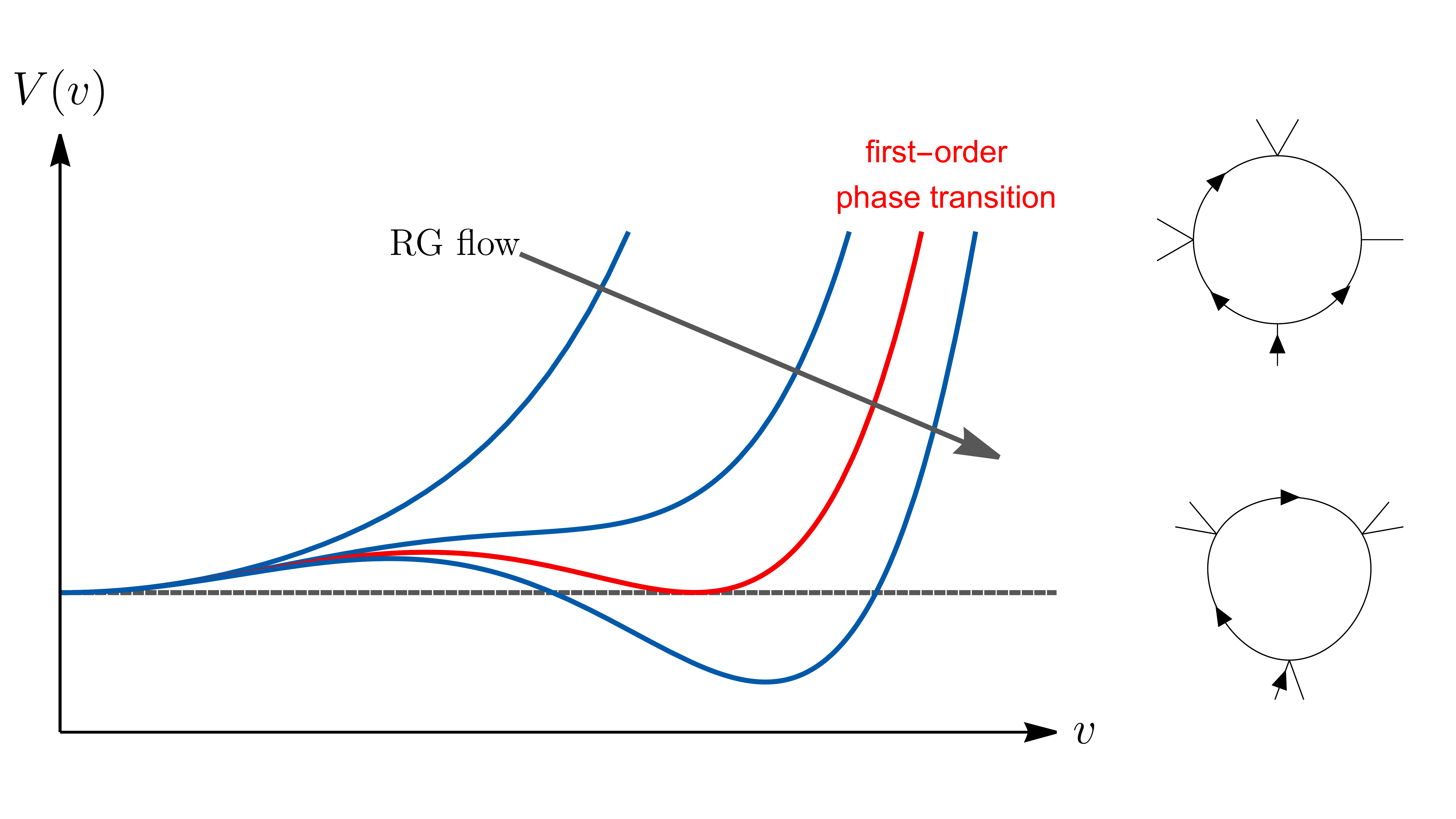}
	\caption{Left - An idea of how the inclusion of the $ \vv^6$ term in the potential, $V_b(\vv) = \frac 12 m_b \vv^2 + \frac 14 J_b \vv^4+ \frac 16 \mathcal K _ b \vv^6$, may affect the renormalization group flow of the Malthusian Toner-Tu theory.  $J_b$ becoming negative, while both $m_b $ and $\mathcal K_b$ staying positive,  could lead to a first-order phase transition phenomenology. When the RG flow enters the $u_b<0$ region (red curve), the perturbative expansion breaks down. Right - Two of the 39 possible Feynman diagrams contributing to the generation of the coupling $\mathcal K_0 \vv^4 \vv$.}
	\label{fefo}
\end{figure}

 We can solve this apparent contradiction considering that the RG transformation may generate additional couplings that grant the stability of the theory also in the $u_b<0$ region. 
In figure \ref{fefo} (right) we show two Feynman diagrams that generate a $\vv^4 \vv $ term in the equation of motion, which may stabilize the system in the $u_b<0$ region. Even if the coefficient $J_b$ becomes negative
\footnote{ The effective coupling constant $u_b$ is linked to the coupling $J_b$ by the simple relation  $u _0 = J_0 D_0 /\Gamma_0^2$ \eqref{effective_par_def}; therefore $u_b$ and $J_b$ have the same sign.}, a force term $-\mathcal K \vv^4 \vv$ could still guarantee  the stability of the equation of motion, provided that $\mathcal K>0$.
Adding this novel force term in the equation of motion is equivalent to modifying the pseudo-potential $V_{\mathrm{new}}(\vv) $ as follows:
\begin{equation}
	V_{\mathrm{new}}(\vv) = \frac 12 m_0 \vv^2 +\frac 1 4 J_0 \vv^4  +\frac 16  \mathcal K_0 \vv^6 
	\label{vader}
\end{equation}

If $\mathcal K_0$ is zero, the potential $V(\vv)$ develops a non-zero minimum only for $m_0<0$; moreover, these minima are arbitrarily close to zero provided that $m_0$  is small enough. This scenario gives rise to the phenomenology of a second-order phase transition. Conversely, if $\mathcal K_0 >0$, the potential $V$ may develop a minimum which is not close to zero when $J_0$ becomes negative. As shown in figure \ref{fefo}, when the constant $J_0$ becomes negative the potential may develop a global minimum far from $\vv = 0$, provided that both $m_0$ and $\mathcal K_0$ remain positive; \textit{this would result in a discontinuous, first-order, phase transition.}

In the literature, this scenario, where one or more couplings become negative and then run away, is often, but not always \cite{toner82,Dasgupta81}, related to a fluctuation-induced first-order transition. The Heisenberg model with cubic anisotropy and the scalar electrodynamics \cite{amit2005field}  are two paradigmatic examples in which run-away trajectories result in a first-order phase transition.  Non-perturbative techniques are often needed to prove that a run-away RG trajectory results in a first-order phase transition \cite{Oerding2000,amit2005field}, but this is beyond the purposes of this perturbative RG calculation.

The stopping condition of the RG equations is determined by the correlation length $\xi$, or by the system size $L$. Depending on $\xi$, the RG flow may enter or not the unstable region. If the correlation length, or the system size, is small the RG flow does not enter the unstable region, and the system displays the phenomenology of a second-order phase transition, ruled by the incompressible active matter fixed point. 
Conversely, if the correlation length is large enough, the RG flow enters the $u_b<0$ region; when it happens because of the strong fluctuations, a phase transition, that should be second-order, becomes first-order \cite{amit2005field}. In this sense, we call this phase transition a fluctuation-induced first-order phase transition \cite{amit2005field}.

\textit{Conclusions - } 
We computed the renormalization group flow for the Malthusian Toner-Tu theory. In the incompressible limit this calculation reproduces the results found in incompressible active matter \cite{chen2015critical}. Thus, in agreement with the numerical evidences \cite{cavagna21equilibrium},  the critical exponents found in \cite{chen2015critical}  hold also for systems with mild density fluctuations, provided that the correlation length (or the system size) is not too large. Furthermore, we found a relation between two \textit{measurable} parameters, $\mu$ and the correlation length $\xi$ (or the size of the system $L$) that determines whether or not the critical dynamics is ruled by the incompressible active matter fixed point. If the system size is large enough, the RG flow escapes the incompressible active matter fixed point, and enters an unstable region. \textcolor{black}{This suggests that, for large enough systems size $L$ and correlation length $\xi$, a fluctuation-induced first-order phase transition may occur}\textcolor{black}{; corroborating the idea that fluctuations and renormalization of parameters play a crucial role in determining the order of the phase transition \cite{Martin2021FIPS}}.	
We believe that even though, in MTT theory, density fluctuations are partially suppressed, they are still strong enough to turn the second-order phase transition into a first-order phase transition.
Further studies are needed to clarify the situation; two strategies are including other couplings in the equation of motion or using non-perturbative techniques. This calculation provides one of the first RG evidence that the phase transition in active matter is first-order for large enough systems. 

We thank A. Cavagna, I Giardina, and T. Grigera for discussions. We thank J. Toner and A. Maitra for comments. This work was supported by ERC grant RG.BIO (Grant No. 785932). L. Di Carlo designed the study, L. Di Carlo and M. Scandolo performed the RG computation. 
\bibliographystyle{apsrev4-2}
\bibliography{Bibbi}
\end{document}


\title{Supplemental material:\\ Evidence of fluctuation-induced first-order phase transition in active matter}
		\author{
	Luca Di Carlo $^{1,2}$
	}
	\begin{abstract}
	\end{abstract}
	\author{
		Mattia Scandolo$^{1,2}$
	}
	\affiliation{$^1$ Dipartimento di Fisica, Universit\`a\ la Sapienza, 00185 Rome, Italy}
		\affiliation{$^2$ Istituto Sistemi Complessi, Consiglio Nazionale delle Ricerche, UOS Sapienza, 00185 Rome, Italy}
	\maketitle
	
    This supplemental material is intended mainly for the renormalization group's technicians interested in the technical details of the computation. In the first part of this SM, we derive the Martin-Siggia-Rose/Janssen-De Dominicis action associated with the equation of motion of the Malthusian Toner-Tu model. We define the essential elements of perturbation theory, namely Gaussian correlation functions and Feynman vertices. The second part discusses the renormalization group calculation and, in particular, we explicitly show all the perturbative contributions that arise from the momentum shell renormalization group procedure. 	In the third part, we discuss the fixed point structure of the theory. In particular, we present several consistency checks that the recursion relations of the Malthusian Toner-Tu model must pass.
	\section{1 - Field theory formalism}
As anticipated, we are using the Martin-Siggia-Rose/Janssen-De Dominicis technique \cite{cardy1996scaling}, a standard technique that allows writing a stochastic differential equation as a field theory, formulated using path integrals. This path integral representation will be the starting point of the renormalization group computation. 

	A stochastic differential equation, determined by a deterministic evolution operator  $\mathcal F$ and a Gaussian noise $\boldsymbol \theta$,
	\begin{CEquation}
		\boldsymbol{\mathcal{F}}\left[\vv\right]-\boldsymbol{\theta}=0 \ \ \mbox{,}
		\label{manbearpig}
	\end{CEquation}
	can be described through a field-theoretical action  $\mathcal S $ that correctly reproduces its statistics, i.e. its the correlation and response functions, 
	\begin{CEquation}
		\mathcal{S}[\boldsymbol{\hat{v}},\boldsymbol{v}]=\int \left[\hat{v}_\alpha \mathcal{F}_\alpha\left[\boldsymbol{v}\right] - \hat{v}_\alpha D_{\alpha\beta} \hat{v}_\beta\right] \ d \boldsymbol{x} dt \qquad ,
		\label{eq:MSRaction}
	\end{CEquation}
	where $2 D_{\alpha\beta}$ is the variance of the Gaussian noise.  The field $\hat {\vv}$ is an auxiliary field, and intuitively it represents a Lagrange multiplier that allows selecting only the field configurations $\vv(\vx,t)$ that satisfy the equation of motion \eqref{manbearpig}; $\hat \vv (x,t)$ is also called the \textit{response field}, since the response function $G(\tvk)$ can be written as \cite{de2006random}:
	\begin{CEquation}
\langle\hat{\boldsymbol{v}}(-\tvk)\boldsymbol {v} (\tvk^\prime)\rangle	=	G(\tvk) \tilde\delta(\tvk-\tvk^\prime) \quad ,
		\label{eq:response}
	\end{CEquation}
where $\tvk$ and  $\tilde \delta(\tvk)$ are defined as follows:
\begin{CEquation}
	\tilde {\boldsymbol{k}} = (\vk, \omega_k) \qquad \qquad \qquad \tilde \delta(\tvk) = (2\pi)^{d+1} \delta^{d}(\vk) \delta(\omega_k)
\end{CEquation} 
The average $\langle \bullet \rangle$ is performed over the pseudo probability distribution $\mathcal P[ \hat {\boldsymbol{v}},\boldsymbol{v}]$:
\begin{CEquation}
	\mathcal P [\hat {\boldsymbol{v}},\boldsymbol{v}] = e^{-\mathcal S [\hat {\boldsymbol{v}},\boldsymbol{v}]}
\end{CEquation}

Applying this procedure to the equation of motion of the Malthusian Toner-Tu theory,
\begin{CAlign}
&\de_ t \vv + \gamma_1 (\vv \cdot \nabla) \vv = -(m_0 + J_0 \vv^2) \vv + \Gamma_0 \nabla^2 \vv  +\gamma_2 (\nabla \cdot \vv) \vv + \gamma_3 \nabla v^2+\sigma_1\frac{\rho_0}{|g^\prime ( \rho_0)|} \nabla( \nabla \cdot \vv) + \boldsymbol f \ , 
		\label{babbala}  
		\\ 
	&	\langle f_{\alpha}(\boldsymbol x,t) f_\beta (\boldsymbol x^\prime t^\prime)\rangle = 2 D_0 \delta_{\alpha \beta} \delta^d(\boldsymbol x - \boldsymbol x^\prime) \delta (t-t^\prime)
\end{CAlign}
leads to the following MSR action $\mathcal S = \mathcal S_0 + \mathcal S_I$,
\begin{CEquation}
	\begin{split} 
		\mathcal S_0[\boldsymbol v, \hat {\boldsymbol{v}}] &= \int d \tvk \ \hat v_{\alpha}(-\tvk)\left [ (- i \omega+m_0 + \Gamma_0 ^\parallel \vk^2)P^{\parallel}_{\alpha \beta} + (-i \omega + m_0 + \vk^2 \Gamma_0 ^\perp) P^{\perp} _{\alpha \beta} \right] v_\beta(\tvk)- D_0 \hat v_\alpha(-\tvk) \hat v_\alpha(\tvk)  \\
		\mathcal S_I [\boldsymbol v, \hat {\boldsymbol{v}}] &= \int d \tvk \  d \tvq  \ i\, \hat v_{\alpha} (-\tvk) \left[ \gamma_{1,0} \delta_{\alpha \beta} q_\gamma + \gamma_{2,0} \delta _{\alpha \gamma} q_\beta +\gamma_{3,0} \delta _{\beta \gamma} k_\alpha  \right] v_\beta(\tvq) v_\gamma (\tvk-\tvq) \\ 
		&	- J_0  \int d \tvk \ d \tvq \ d \tvp \  \hat v_{\alpha} (-\tvk) Q_{\alpha \beta \gamma \nu}  v_\beta (\tvq) v_\gamma (\tvp) v_\nu(\tvk-\tvq-\tvp) 
	\end{split} 
	\label{MS_eq:action} 
\end{CEquation}
where we have defined the longitudinal and transverse projection operators, and the longitudinal and transverse kinetic coefficients
\begin{CAlign}
&P^\perp_{\alpha \beta}(\boldsymbol k) = \delta_{\alpha \beta} - {k_\alpha k _\beta} /{k^2} \qquad , \quad P^ {\parallel}_{\alpha \beta} (\boldsymbol k) = {k_\alpha k_\beta} /{k^2} \  \mbox{,}
\label{potte}
 \\
&\Gamma^\perp _0 = \Gamma_0 \qquad\qquad\qquad  \quad \qquad, \quad \Gamma_0 ^\parallel = \Gamma_0 + \sigma_1 \frac{\rho_0}{|g^\prime(\rho_0)|}
\label{duccio} 
\end{CAlign}

We distinguish between the Gaussian and non-Gaussian parts of the action: $\mathcal  S _0$ is the Gaussian action, corresponding to the linear terms in the equation of motion \eqref{babbala}, and $\mathcal S_I$ is the non-Gaussian action, corresponding to the non-linear terms in the equation of motion \eqref{babbala}. 

We remark that the action has the standard structure shown in equation \eqref{manbearpig}: the linear part in the auxiliary field $\hat {\vv}$ multiplies the deterministic part of the equation of motion, while the quadratic term in the auxiliary field $\hat {\vv}$ multiplies the noise amplitude.
\subsection{Linear correlation functions} 
The Gaussian two-point functions represent the building blocks of the perturbative expansion. In the Malthusian Toner-Tu theory, because of the anisotropy given by the pressure force, the two-point functions are not diagonal in the spatial indices, which means that $\langle v_\alpha v_\beta \rangle$ is not proportional to $\delta_{\alpha\beta}$. To derive the  two-point function, we write the Gaussian action $\mathcal S_0$ in a matricial form:

\begin{CEquation}
	\mathcal S_0 = \frac 12 \int d \tvk \phi_\alpha(\tvk) \mathbb  M_{\alpha \beta}(\tvk) \phi_\beta(-\tvk)\qquad \qquad \mathbb M_{\alpha\beta}(\tvk) = 
	\begin{pmatrix} 
		0 & A^\perp P^\perp_{\alpha\beta} + A^\parallel P^\parallel _{\alpha \beta}  \\ 
\bar A^\perp P^\perp_{\alpha\beta} +\bar A^\parallel P^\parallel _{\alpha \beta} &-2D_0 \delta_{\alpha\beta}
	\end{pmatrix} \quad , 
\label{cartman}
\end{CEquation}
where the vector $\boldsymbol\phi$ is defined as $\boldsymbol\phi(\tvk) = (\vv(\tvk), \hat \vv(\tvk))$, while $A^\perp$ and $A^\parallel$ are defined as follows
\begin{CEquation}
	A^{\perp,\parallel} = -i \omega + m_0+\Gamma_0 ^{\perp,\parallel} \vk^2 \mbox{,}
\end{CEquation}
and $\bar A^{\parallel,\perp}$ are the respective complex conjugate. We remark that the matrix $\mathbb M$ is a $2d \times 2d$ matrix, and each block of the matrix, as defined in equation \eqref{cartman}, is a $d\times d$ block.

Correlation and response functions are obtained by inverting the matrix $\mathbb M_{\alpha\beta}(\tvk)$.  The matrix $\mathbb M_{\alpha \beta}$ is a block matrix, and also its inverse is a block matrix:
\begin{CEquation}
 \mathbb G_{\alpha\gamma}(\tvk) \mathbb  M_{\gamma \beta}(\tvk) = \begin{pmatrix}
 	\delta _{\alpha\beta} &0 \\ 
 	0 &\delta_{\alpha\beta} 
 	\end{pmatrix}	\qquad \quad \mathbb G_{\alpha\beta} = \begin{pmatrix}
 	C_{\alpha\beta} &\bar G_{\alpha\beta} \\ 
G_{\alpha\beta} &0 
 \end{pmatrix} \qquad ,
\label{mollo}
\end{CEquation}
The matrix elements of $\mathbb G_{\alpha \beta} (\tvk)$ are the Gaussian correlation and response function \cite{parisi_book,cardy1996scaling}; $C_{\alpha\beta}$ is the Gaussian correlation function, and  $G_{\alpha\beta}$ is the Gaussian response function, 

\begin{CEquation} 
\langle \hat v_{\alpha}(-\tvk) v_{\beta}(\tvk')  \rangle = G_{\alpha\beta} (\tvk) \tilde \delta(\tvk-\tvk') \qquad \qquad \qquad \langle v_{\alpha}(-\tvk) v_{\beta}(\tvk')  \rangle = C_{\alpha\beta} (\tvk) \tilde \delta(\tvk-\tvk') \qquad . 
\end{CEquation}
Expanding equation \eqref{mollo} gives a  matricial equations for $G_{\alpha\beta}$ and $C_{\alpha\beta}$, 
\begin{CEquation}
	G_{\alpha\gamma} (A^\perp P^\perp_{\gamma \beta} + A^\parallel P^\parallel _{\gamma\beta }) = \delta_{\alpha\beta} \qquad \quad C_{\alpha\gamma} (\bar A^\perp P^\perp_{\gamma\beta} +\bar A^\parallel P^{\parallel}_{\gamma\beta}) -2 D_0 G_{\alpha\beta} = 0
\end{CEquation}
that can be solved making the following anstaz,
\begin{CEquation}
	\begin{split}
		& G_{\alpha \beta}(\vk, \omega) = P_{\alpha \beta}^{\parallel} (\vk) G^{\parallel}_0 (\vk, \omega) + P_{\alpha \beta} ^{\perp} (\vk) G_0 ^{\perp} (\vk,\omega) \\ 	& C_{\alpha \beta}(\vk, \omega) = P_{\alpha \beta}^{\parallel} (\vk) C^{\parallel}_0 (\vk, \omega) + P_{\alpha \beta} ^{\perp} (\vk) C_0 ^{\perp} (\vk,\omega) \qquad .
	\end{split} 
\label{Corfu}
\end{CEquation}
through which we can easily find the following solution:
\begin{CEquation}
	G_0 ^ {\perp, \parallel}(\vk,\omega) = \frac {1}{- i \omega + m_0 +\Gamma_0 ^{\perp, \parallel} k^2} \qquad 	C_0 ^ {\perp, \parallel}  (\vk,\omega)= \frac {2 D_0}{\omega^2 + ( m_0 + \Gamma_0 ^{\perp,\parallel} k^2)^2} 
\end{CEquation}
As made explicit by equation \eqref{Corfu} the transverse and longitudinal modes are non-interacting in the linear theory. This means that the longitudinal and transverse fluctuations of the velocity field are independent at the Gaussian level.

\subsection{Non-linear terms}
Of course, even if the longitudinal and transverse fluctuations are not coupled at the Gaussian level, they interact at the non-linear level. There are four non-linearities, respectively $\gamma_{1,0}$, $\gamma_{2,0}$, $\gamma_{3,0}$, and $J_0$, which mix the longitudinal and transverse components of the velocity field.
In principle, each Feynman vertex represents a non-linearity, however
since the first three non-linearities involve two fields $\boldsymbol{v}$ and one field $\hat{ \boldsymbol v}$, we can represent them with only one Feynman vertex. Therefore, the theory has only two Feynman vertices, defined as follows:
\begin{CAlign} 
	\begin{gathered} 
		\includegraphics[]{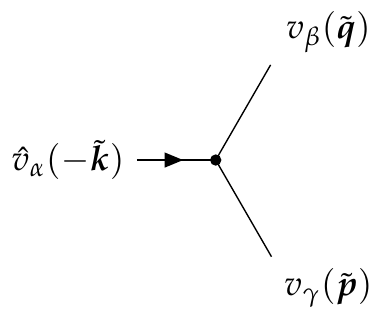}
	\end{gathered} & \qquad =  \ - \frac i 2  Y_{\alpha \beta \gamma} (\vk,\vq,\vp) \tilde \delta( \tvk- \tvq-\tvp)\qquad ,   \\ 
	\qquad &  
	\label{MS_eq:VertexSpp} 
	\\ 
	\begin{gathered} 
		\includegraphics[]{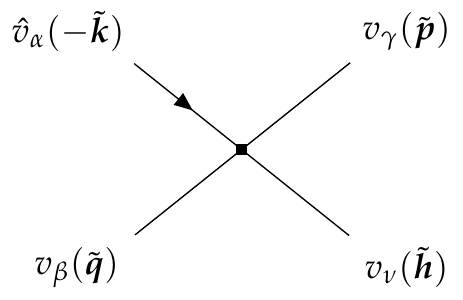} 
	\end{gathered}  & \qquad =\  -\frac{ J_0}{3} Q_{\alpha \beta \gamma \nu} \tilde  \delta( \tvk-  \tvq- \tvp-\tvh) \qquad ,
	\label{MS_eq:VErtexFerro}
\end{CAlign} 
where the tensors  $Y_{\alpha \beta \gamma} ( \vk,\vq,\vp)$ and $Q_{\alpha \beta \gamma \nu}$ are defined as follows: 
\begin{CAlign} 
	& Y_{\alpha \beta \gamma} (\vk, \vq, \vp) =  \gamma_1  (\delta_{\alpha \gamma} p_\beta + \delta_{\alpha \beta} q_\gamma) + \gamma_2 ( \delta_{\alpha \gamma} q_\beta + \delta_{\alpha \beta} p_\gamma ) + 2 \gamma_3 \delta_{\beta \gamma} k_\alpha 
	\label{MS_eq:Ydef}
	\\
	& Q_{\alpha \beta \gamma \nu} =   \delta_{\alpha \beta}\delta _{\gamma \nu}  + \delta_{\alpha \gamma} \delta_{\beta \nu} + \delta_{\alpha \nu} \delta_{\beta \gamma}
	\label{MS_eq:Qdef}
\end{CAlign}
We remark that we symmetrized each non-linearity, in order for each interaction term to be symmetric under exchange of the velocity fields.
We call the first vertex \eqref{MS_eq:VertexSpp} the self propulsion vertex, since it arise from the particle being self propelled at the microscopic level. The self propulsion vertex is composed by the convective derivative term, proportional to $\gamma_{1,0}$, and by the two terms, $\gamma_{2,0}$ and $\gamma_{3,0}$, which are introduced because of the Galilean-invariance breaking. The second vertex \eqref{MS_eq:VErtexFerro} is the standard "ferromagnetic" relaxation vertex, characteristic of ferromagnetic theories belonging to the $\mathcal O(n)$ universality class \cite{parisi_book,kardar2007statistical}.

\section{2 -  Renormalization group equations} 
The momentum-shell renormalization group unfolds through two stages.
The first stage consists of integrating out large momenta fluctuations, precisely fluctuations inside the momentum shell $\Lambda/b <k<\Lambda$. The key point is that $b$, which determines how many modes we are integrating out, is close to one, $b \simeq 1$; this means that we are integrating out a thin shell of modes, defining in this way a continuous transformation. After the integration, we are left with a theory whose cut-off is not $\Lambda$ anymore, but it is $\Lambda/b$.

The second step of the renormalization group consists of  the following rescaling of momenta, frequencies and fields,
 \begin{CEquation} 
	\begin{split}
		&\omega^\prime = \omega b^{z} \qquad \qquad \qquad \hat v^\prime (k,\omega) = b^{-d_{\hat v}} \hat v( b^{-1} k,b^{-z}\omega) \\ 
		& k^\prime = k b \qquad \qquad \qquad
		\ v^\prime (k,\omega) = b^{-d_{ v}}  v( b^{-1} k,b^{-z}\omega) \ \ , 
	\end{split}
	\label{MS_eq:scaling_tranformation} 
\end{CEquation} 
whose effect is to restore the cut-off to its original value $\Lambda$.
Here the exponent $z$ is called the \textit{dynamical exponent} and it rules the interplay between spatial and temporal scales. 

If we apply these two steps to the action of the  Malthusian Toner-Tu model \eqref{MS_eq:action} we end up with the following action,
\begin{CEquation}
	\begin{split} 
		\mathcal S_{0,b}[\boldsymbol v, \hat {\boldsymbol{v}}] &= \int d \tvk \ b^{d_v+d_{\hat v} -d-z} \ \hat v_{\alpha}(-\tvk)\left [ (- i b^{-z}\omega+m_0 + b^{-2}\vk^2\Gamma_0 ^\parallel )P^{\parallel}_{\alpha \beta} + (-ib^{-z} \omega + m_0 + b^{-2}\vk^2 \Gamma_0 ^\perp) P^{\perp} _{\alpha \beta} +\Sigma_{\alpha\beta} \right] v_\beta(\tvk) \\ 
		&  \qquad \qquad \qquad \qquad \qquad \quad - b^{2 d_{\hat v} -d-z} \left[D_0 \delta_{\alpha\beta} +\tilde \Sigma_{\alpha\beta}\right] \hat v_\alpha(-\tvk) \hat v_\beta(\tvk)  \\
		\mathcal S_{I,b} [\boldsymbol v, \hat {\boldsymbol{v}}] &= - \int d \tvk \  d \tvq \ b^{d_{\hat v} + 2 d_v -2d-2z-1}\  \hat v_{\alpha} (-\tvk) \left( Y_{\alpha\beta\gamma} + V^{\hat v v v } _{\alpha\beta\gamma} \right) v_\beta(\tvq) v_\gamma (\tvk-\tvq) \\ 
		&	- J_0  \int d \tvk \ d \tvq \ d \tvp \ b^{d_{\hat v} + 3 d_v -3d-3z } \hat v_{\alpha} (-\tvk) (Q_{\alpha \beta \gamma \nu} + V^{\hat v v v v}_{\alpha\beta\gamma\nu} )   v_\beta (\tvq) v_\gamma (\tvp) v_\nu(\tvk-\tvq-\tvp)  \qquad ,
	\end{split} 
	\label{algore} 
\end{CEquation}
where the subscript $b$ indicates that this is the action after the renormalization group transformation. 
We remark that the structure of this novel action is the same as the original one. Higher order terms may be generated, but they are irrelevant at the first order in  $\epsilon$ \cite{Wilson1974}.
The action after the RG transformation differs from the original one because of two different kinds of contributions: naive rescaling and perturbative contributions. The perturbative contributions are given by the momentum shell integral, and they are determined by the terms  $\Sigma_{\alpha\beta}$, $\tilde \Sigma_{\alpha\beta}$, $V^{\hat v v v }_{\alpha\beta \gamma} $ and $V^{\hat v v v v}_{\alpha\beta\gamma\nu} $.  The actions \eqref{MS_eq:action} and \eqref{algore} have the same structure, but their parameters are different. The parameters after and before the RG transformation are linked by the following recursion relations,
\begin{CAlign}
	& m_b = m_0 \  b^{z} (1+\delta m \ln b ) 
\label{b1}
\\
	& \Gamma^\perp_b = \Gamma_0^\perp \  b^{z-2 -\eta} (1+\delta \Gamma^\perp \ln b ) 
	\label{b21}
	\\
	& \Gamma^\parallel = \Gamma_0^\parallel \  b^{z-2 -\eta} (1+\delta \Gamma^\parallel \ln b ) \\
	&\gamma_{i,b} = \gamma_{i,0} \ b^{d_{\hat v}+2d_v-2d-2z-1} (1 + \delta\gamma_i \ln b) \\
	&  J_b = J_0 \ b^{d_{\hat v} +3d_v -3z-3d}(1+\delta J \ln b) \\ 
	& D_b = D _0 \ b^{2 d_{\hat v} -d-z}(1+ \delta D \ln b)
	\label{b2}
\end{CAlign}
where we denote with $X_b$ the parameter $X$ after the RG transformation, and with $\delta X$ the perturbative contribution to the parameter $X$.  
We explicitly highlighted the fact that perturbative corrections are proportional to $\ln b$.
Finally we impose a relation between the scaling dimensions of the fields $\vv$ and $\hat \vv$,
\begin{CEquation}
	d_{\hat v}  = -d_v +d+2z+ \eta \ \ \mbox{,}
	\label{giggio}
\end{CEquation}
which is obtained by imposing that the term $-i\omega$ of the action \eqref{algore}  remains unchanged after the renormalization group transformation. Here, $\eta$ is an anomalous dimension contribution, and it is determined by the term of order $i \omega$ in the self-energy $\Sigma_{\alpha\beta}(k,\omega)$. Plugging the expression for $d_{\hat v}$ in equations  \eqref{b1}-\eqref{b2}, we obtain equations (17)-(21) of the main text.

In order to determine the RG perturbative contributions we have to compute four different vertex functions: two self energies (two point functions) $\Sigma(k,\omega)$ and $\tilde \Sigma(k,\omega)$, the $V^{\hat v v v }$ three point function and the $V^{\hat v v v v} $ four point function.
\begin{itemize}
	\item The self energy $\Sigma_{\alpha \beta} $ corrects the mass term $m_0$, the kinetic coefficients $\Gamma_0^ \parallel$, $\Gamma_0 ^\perp$, and it will eventually provide an anomalous dimension correction $\eta$;
	\item  the self energy $\tilde \Sigma_{\alpha \beta}$ corrects the noise amplitude $D_0$;
	\item the three point function $V^{\hat v v v} _{\alpha \beta \gamma} $ corrects the couplings $\gamma_{1,0}$, $\gamma_{2,0}$ and $\gamma_{3,0,}$; 
	\item the four point function $V^{\hat v v v v }_{\alpha \beta \gamma \nu}  $ corrects the ferromagnetic coupling $J_0$.
\end{itemize}
Before proceeding we define a set of effective parameters and effective couplings, whose scaling dimension is independent of the scaling dimension of the field and on $z$:
\begin{CEquation}
	\mu_0 = \frac{\Gamma^\parallel _0}{\Gamma^\perp _0}  \qquad \quad 
  	  u_0 = \frac{J_0}{ \Gamma_0 ^\perp} \frac{D_0} {\Gamma_ 0 ^{\perp}} \Lambda^{4-d}  K_d \quad \qquad 
	g_{i,0} = \frac{\gamma_{i,0} }{\Gamma_0^\perp }  \sqrt{\frac{D_0}{\Gamma_0 ^\perp}}\sqrt{\Lambda^{4-d} K_d}
	\label{MS_eq:effectiveParDef} \quad \qquad r_0 = \frac{m_0} {\Gamma_0 ^\perp}
\end{CEquation}
where $\Lambda$ is the cutoff of the theory and $K_d = S_d/(2\pi)^d$, being $S_d$ the surface of the $d$-dimensional hyper-sphere.  It is useful to describe the renormalization group flow in terms of these effective parameters; first, the perturbative contributions can be written in terms of these effective parameters; furthermore, all the critical exponents are fully determined by the renormalization group flow of these effective parameters. 
We can obtain the recursion relations of the effective parameters by the recursion relations \eqref{b1}-\eqref{b2}:
\begin{CAlign}
	r_b & = b^2 r_0 \left[ 1 + (\delta m - \delta \Gamma^\perp) \ln b \right] \label{MS_eq:recursionR}\\
	\mu_b &= \mu_0 \left[1+ (\delta \Gamma^\parallel  - \delta \Gamma^\perp )\ln b\right] \label{MS_eq:recursionU} \\
	u_b &= b^{\epsilon} u_0 \left[ 1 + (\delta J + \delta D -2 \delta \Gamma^\perp -\eta) \ln b \right]\\
	g_{1,b} &  = g_{1,0} b^{\epsilon /2} \left[1+ \left (\delta \gamma_1 + \frac 12 \delta D- \frac 32 \delta \Gamma^\perp-\eta\right) \ln b\right]\\
	g_{2,b} &  = g_{2,0} b^{\epsilon /2} \left[1+ \left (\delta \gamma_2+ \frac 12 \delta D- \frac 32 \delta \Gamma^\perp-\eta\right) \ln b\right]\\
	g_{3,b} &  = g_{3,0} b^{\epsilon /2} \left[1+ \left (\delta \gamma_3 + \frac 12 \delta D- \frac 32 \delta \Gamma^\perp-\eta\right) \ln b\right] \label{MS_eq:recursiong3}
\end{CAlign}
The naive scaling dimensions of the effective parameters are:
\begin{CEquation}
	d_\mu = 0 \qquad\qquad  \qquad d_{g_i} = \epsilon/2 \qquad \qquad\qquad  d_u = \epsilon \qquad \qquad \qquad d_r = 2
	\label{MS_eq:effectiveParGScaling} 
\end{CEquation}
The effective coupling constant $u$ has naive scaling dimension  $\epsilon$, suggesting that its fixed point value must also be of order $\epsilon$. Similarly, since the coupling constants $g_i$ have naive scaling dimension $\epsilon /2$, their fixed point will be of order $\epsilon ^{1/2}$.
The parameter $r$ represents the distance from the transition point, and it is analogous to the distance from the critical temperature $T-T_c$ in equilibrium systems. At the mean-field level, the transition point corresponds to $r = 0$; however, because of the non-linear terms, the transition point is shifted of $\mathcal O(g_i^2,u)$ from its naive value.
We conclude that also the critical value of $r$ must be of the same order $r_c = \mathcal O (\epsilon)$. Since we are performing an expansion in powers of $\epsilon$, we can neglect, at the leading order in $\epsilon$, the $r$ dependence in most of the Feynman diagrams, since this would lead to corrections of order $\epsilon ^2$.

\subsection{ Perturbative contributions} 
The two point functions $\Sigma_{\alpha\beta}$ and $\tilde \Sigma_{\alpha\beta}$ are given by the following Feynman diagrams expansion:
\begin{CEquation}
	  \Sigma_{\alpha \beta} (\vk,\omega) = 
	\begin{gathered} 
		\includegraphics[trim={3cm 0 0 0},clip]{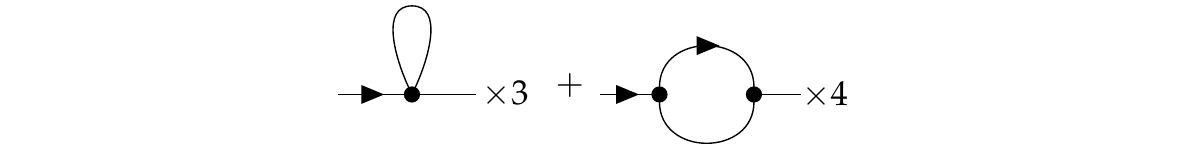} 
	\end{gathered}\hspace{-1.7cm}
\tilde \Sigma_{\alpha \beta}(\vk,\omega)  = 
\begin{gathered} 
		\includegraphics[trim = {4.5cm 0 0 0}]{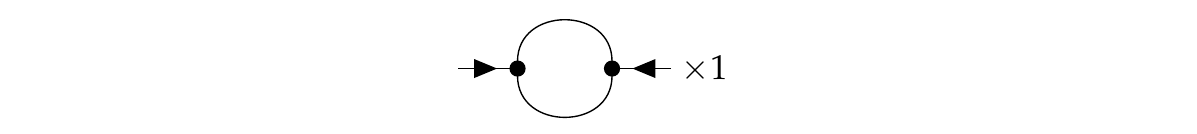}
	\end{gathered}  
\end{CEquation}
The self-energy $\tilde \Sigma_{\alpha \beta}$ corrects only the noise amplitude $D_0$, and for this reason, it is sufficient to compute it at the zeroth order in the external momenta:
\begin{CEquation}
	\tilde \Sigma_{\alpha \beta}  =  D \frac{3 ({g_1}+{g_2})^2}{2 \mu  (\mu +1)}  \delta_{\alpha \beta}  \ln b \qquad \qquad \delta D = \frac{3 ({g_1}+{g_2})^2}{2 \mu  (\mu +1)}
\end{CEquation}

The structure of the self energy $\Sigma_{\alpha \beta} (\vk,\omega) $ is more complicated, and it corrects many parameters. $\Sigma_{\alpha \beta}(\vk,\omega)$ contains  isotropic, longitudinal and transverse terms respectively proportional to $\delta_{\alpha \beta}$, $P^\parallel _{\alpha \beta} (\vk)$ and to $P^{\perp} _{\alpha \beta} (\vk)$. We write directly the expression of the self energy isolating the various contributions: 
\begin{CEquation} 
	\Sigma_{\alpha \beta} (k,\omega) =( m_0 \delta m -i \omega \eta)\delta_{\alpha \beta} \ln b + \left[  \Gamma_0 ^\parallel  k^2  \ \delta \Gamma ^\parallel P^{\parallel}_{\alpha \beta} (\vk) + \Gamma^{\perp} _0 k^2 \ \delta \Gamma ^\perp  P^{\perp} _{\alpha \beta} ( \vk) \right] \ln b  \qquad ,  
	\label{MS_eq:Dgammatilde}
\end{CEquation}
where
\begin{CAlign}
	&\delta m  = \frac{3 u (3 \mu +4 r+1)}{2 r (r+1) (\mu +r)}-\frac{3 \left(g_1+g_2\right) \left(g_2 (r+1)-2 g_3 (\mu +r)\right)}{4 r
		(r+1) (\mu +r) (\mu +2 r+1)}
	\label{dm}
	\\ 
	&\eta =\frac{3 ({g_1}+{g_2}) ({g_2}-2 {g_3} \mu )}{4 \mu  (\mu +1)^2}
	\label{eta}
\end{CAlign}
\begin{CEquation}
	\begin{split} 
		\delta \Gamma^{\perp} =&
		g_1^2 \frac{3 \mu^3+8 \mu^2+10\mu-3}{12\mu (\mu+1)^2}-
		g_1 g_2\frac{2 \mu ^4-12 \mu ^2+\mu +1}{12 \mu ^2 (\mu +1)^3} + \\
		+& g_2 g_3 \frac{12 \mu ^2+3 \mu
		+1}{6 \mu ^2 (\mu +1)^3}-g_1 g_3 \frac{5  \mu  (\mu +3)}{6 (\mu +1)^3}-g_2^2\frac{ \mu -9}{12 (\mu +1)^3}
	\end{split} 
	\label{MS_eq:Dgammaperp}
\end{CEquation}
\begin{CEquation}
	\begin{split} 
		\delta \Gamma^\parallel = &
		g_1^2\frac{ 3\mu^2+8\mu+1}{8 \mu  (\mu +1)^2}-
	    g_1 g_2 \frac{6 \mu ^5+14 \mu ^4+9 \mu ^3-3 \mu ^2+7 \mu +7}{8 \mu ^3 (\mu +1)^3}\\
	    -& g_1 g_3 \frac{4 \mu ^2+9 \mu +3}{2 \mu  (\mu +1)^3}+g_2^2\frac{6 \mu ^4+17 \mu ^3+15 \mu ^2-3 \mu -3}{8 \mu ^3 (\mu +1)^3}\\
	    +& g_2 g_3 \frac{12 \mu ^5+36 \mu ^4+39 \mu ^3+27 \mu ^2+15 \mu +7}{4 \mu ^3 (\mu +1)^3}+
		g_3^2\frac{3}{\mu  (\mu +1)}
	\end{split} 
	\label{MS_eq:Dgammalon}
\end{CEquation}
We omitted the subscript zero in all the constants for easier reading. We recall that we must keep the $r_0$ dependency only in the corrections of the mass term $m$.

The vertex function $V^{\hat v v v }_{\alpha \beta \gamma} ( \vk,\vq,\vp)$ is given by the following Feynman diagrams expansion,  
\begin{CEquation}
	V^{\hat v v v } _{\alpha \beta \gamma}  = 
	\begin{gathered}
		\includegraphics[]{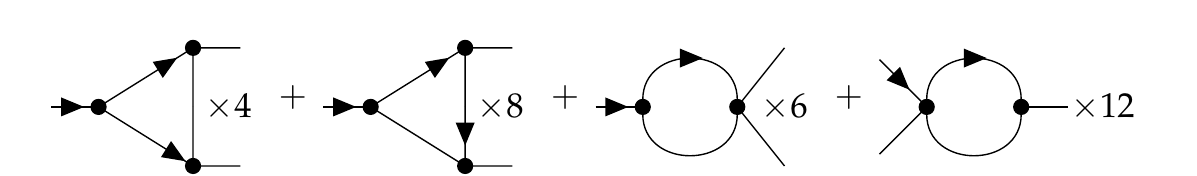} 
	\end{gathered} \qquad , 
\end{CEquation}
and it corrects the couplings $\gamma_{1,0}$, $\gamma_{2,0}$ and $\gamma_{3,0}$; therefore we must compute these Feynman diagrams up to the linear order in the external momenta. To be as clear as possible, we show explicitly how this term corrects the action:
	\begin{CEquation}
		\delta \mathcal S = - \frac i 2 \int \hat  v_{\alpha} (-\tvk)V_{\alpha \beta \gamma}^{\hat v v v} (\vk,\vq,\vp) v_\beta (\tvp) v_\gamma (\tvp) \delta (\tvk -\tvq-\tvp)
		\label{MS_eq:sppActionCorrection} 
	\end{CEquation}
	
Computing explicitly the Feynman diagrams leads to the following expression, 
	\begin{CEquation}
		V^{\hat v v v } _{\alpha \beta \gamma}  (\vk,\vq,\vp)  = 
		\gamma_1  \delta \gamma_1 (\delta_{\alpha \gamma} p_\beta + \delta_{\alpha \beta} q_\gamma) + \gamma_2\delta \gamma_2 ( \delta_{\alpha \gamma} q_\beta + \delta_{\alpha \beta} p_\gamma ) + 2 \gamma_3 \delta \gamma_3 \delta_{\beta \gamma} k_\alpha  \qquad , 
		\label{MS_eq:SppVertex} 
	\end{CEquation}
where 
	\begin{CEquation}
			\begin{aligned} 
				\delta \gamma_1 =&
				g_1 g_2 \frac{\mu ^2+9 \mu -1}{12 \mu ^2 (\mu +1)^2}
				-g_1 g_3 \frac{\mu ^2+5 \mu +1}{2 \mu (\mu +1)^2}
				+ g_2^2 \frac{8 \mu ^3+28 \mu ^2+10 \mu -1}{12 \mu ^2 (\mu +1)^3}\\
				&-g_2 g_3\frac{ \left(10 \mu ^4+31 \mu ^3+9 \mu ^2+5 \mu -1\right)}{6 \mu ^2 (\mu +1)^3}
				+g_3^2 \frac{3}{(\mu +1)^3}\\
				&-\frac{g_2^2 \left(g_2+2 g_3\right)}{g_1}\frac{3}{4 \mu (\mu+1)^3}
				+\frac{g_2 g_3 \left(g_2+2 g_3\right)}{g_1}\frac{5\mu^2+15\mu+1}{6 \mu (\mu+1)^3}\\
				&- u \frac{ 10\mu^3 +13\mu^2+1}{6 \mu ^2 (\mu +1)}
				+ \frac{u \left(g_2+2 g_3\right)}{g_1} \frac{5 \mu ^3+10 \mu ^2+2 \mu+1}{6 \mu ^2 (\mu +1)^2}
			\end{aligned} 
		\label{MS_eq:deltagamma1}
	\end{CEquation}

	\begin{CEquation}
			\begin{aligned}
				\delta \gamma_2 =& -g_1^2\frac{ 2 \mu +1}{24 \mu ^2 (\mu +1)}
				+ g_2^2 \frac{25 \mu ^2+36 \mu +9}{8 \mu ^2 (\mu +1)^3}
				-g_3^2 \frac{2 \mu ^2+6 \mu +1}{3 \mu  (\mu +1)^3}\\
				&-g_1 g_2\frac{8 \mu ^3-17 \mu ^2-41 \mu -13}{12 \mu ^2 (\mu +1)^3}
				-g_1 g_3 \frac{ 7 \mu ^4+52 \mu ^3+54 \mu ^2-4 \mu -1}{12 \mu ^2 (\mu +1)^3}\\
				&-g_2 g_3 \frac{18 \mu ^4+82 \mu ^3+63 \mu ^2-16 \mu -3}{12 \mu ^2 (\mu +1)^3}
				-\frac{g_1^2 g_3}{g_2}  \frac{1}{4 (\mu +1)}
				+\frac{g_1 g_3^2}{g_2} \frac{ \mu }{ (\mu +1)^3}\\
				&+\frac{g_1 u}{g_2} \frac{ 11 \mu ^3+5 \mu ^2-3 \mu -1}{12 \mu ^2 (\mu +1)}
				-\frac{g_3 u}{g_2} \frac{22 \mu ^4+39 \mu ^3+15 \mu ^2-3 \mu -1}{6 \mu ^2 (\mu +1)^3}\\
				&- u \frac{54 \mu ^5+172 \mu ^4+195 \mu ^3+123 \mu ^2+75 \mu +29}{12 \mu ^2 (\mu +1)^3}
			\end{aligned} 
		\label{MS_eq:deltagamma2}
	\end{CEquation}
	
	\begin{CEquation}
			\begin{aligned} 
				\delta \gamma_3 = & g_1^2 \frac{\mu ^2+3 \mu -1}{4 \mu  (\mu +1)^2}
				+g_2^2  \frac{7 \mu ^3+24 \mu ^2+14 \mu +3}{12 \mu ^2 (\mu +1)^3}
				+g_3^2 \frac{3}{\mu  (\mu +1)}
				-g_1 g_3 \frac{4 \mu ^2+6 \mu +3}{2 \mu  (\mu +1)^3}\\
				&+g_1 g_2 \frac{2 \mu ^4+5 \mu ^3+9 \mu ^2+7 \mu +4}{6 \mu ^2 (\mu +1)^3}
				-g_2 g_3 \frac{9 \mu ^4+23 \mu ^3+24 \mu ^2+22 \mu +9}{6 \mu ^2 (\mu +1)^3}\\
				&+\frac{g_1^2 g_2}{g_3} \frac{5 \mu ^2-3 \mu +1}{24 \mu ^2 (\mu +1)^2}
				+\frac{g_1 g_2^2}{g_3} \frac{4 \mu ^3+5 \mu ^2-\mu +1}{24 \mu ^2 (\mu +1)^3}
				+\frac{g_2^3}{g_3} \frac{1}{8 (\mu +1)^3}\\
				&+\frac{u g_1}{g_3}\frac{13 \mu ^3+4 \mu ^2-3 \mu +10}{12 \mu ^2 (\mu +1)}
				-\frac{u g_2}{g_3}\frac{11 \mu ^4+27 \mu ^3+12 \mu ^2-9 \mu -5}{12 \mu ^2 (\mu +1)^3}\\
				&-u \frac{27 \mu ^4+59 \mu ^3+46 \mu ^2+20 \mu +10}{6 \mu ^2 (\mu +1)^2}
			\end{aligned}
		\label{MS_eq:deltagamma3}
	\end{CEquation}
 
 At this point, a remark is in order: the vertex function $V_{\alpha \beta \gamma} ^{\hat v v v} (\vk,\vp,\vp)$ at $\gamma_2 = \gamma_3 = 0$ takes the following form, 
	\begin{CEquation}
		\begin{split}
			V^{\hat v v v } _{\alpha \beta \gamma}  (\vk,\vq,\vp)  \biggl|_   {\gamma_2 = \gamma_3 = 0}= 
			\frac{u \gamma_1 }{12 \mu^2 (1+\mu)}  [-2(10 \mu^3+13\mu^2+1)
			(\delta_{\alpha \gamma} p_\beta + \delta_{\alpha \beta} q_\gamma) \\
			+(11 \mu^3+5 \mu^2-3\mu-1) ( \delta_{\alpha \gamma} q_\beta + \delta_{\alpha \beta} p_\gamma ) + (13 \mu^3+ 4\mu^2-3\mu+10)\delta \gamma_3 \delta_{\beta \gamma} k_\alpha ]
		\end{split}  \ .
		\label{giangiugia}
	\end{CEquation}
This equation contains a term of the form $\delta_{\alpha \gamma} p_\beta$, corresponding to a $\vv\cdot \nabla \vv$ term in the equation of motion,  and two terms of the form $(\delta_{\alpha\gamma}  q_\beta+  \delta_{\alpha\beta} p_\gamma )$ and $\delta_{\beta\gamma} k_\alpha$, which correspond respectively to  the $\nabla \vv^2$ and $\vv \nabla\cdot \vv$ terms in the equation of motion. This means that, if we start the computation with $\gamma_2 = \gamma_3 = 0$, namely without the terms arising from the Galilean invariance breaking, the renormalization group generates them spontaneously. 
	
	Finally, the vertex function $V^{\hat v v v v} _{\alpha \beta\gamma\nu}$ determines the perturbative corrections to the ferromagnetic coupling $J_0$, and it is given by the following Feynman diagrams expansion: 
	\begin{CEquation} 
		\begin{gathered}
			\includegraphics[scale = 1, trim = 0 1.5cm 0 0 ]{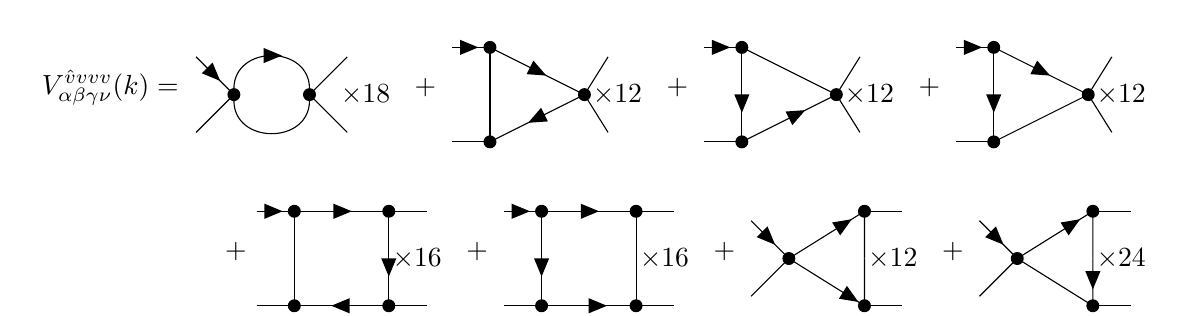}
		\end{gathered}
		\vspace{1.5 cm}
	\end{CEquation} 
We distinguish between three contributions: the bubble diagram, proportional to $J_0^2$; The triangular diagrams, proportional to $J_0 \gamma_i ^2$ and the squared diagrams, proportional to $\gamma_i ^4$. The tensorial structure of the four-point function reproduces the same structure of the ferromagnetic vertex \eqref{MS_eq:VErtexFerro}. This result is quite surprising; the Gaussian theory is anisotropic, and it is not obvious that the renormalization group transformation does not generate anisotropic non-linear terms, namely with a different tensorial structure from the one of equation \eqref{MS_eq:VErtexFerro}. The four-point function $V^{\hat vvvv} _{\alpha \beta \gamma \nu}$ takes the following form:
	\begin{CEquation}
		V^{\hat v v v v } _{\alpha \beta \gamma \nu} = -\frac{J_0}{3} Q_{\alpha \beta \gamma \nu}  \delta J \qquad ,
		\label{MS_eq:Vhvvvv} 
	\end{CEquation}
	where $\delta J$ determines the perturbative corrections to the coupling $J_0$:
	\begin{CEquation}
			\begin{aligned} 
				\delta J =&
				-u \frac{17 \mu ^2+2 \mu +5}{2 \mu ^2}
			    +g_2^2 \frac{7 \mu ^3+32 \mu ^2+29 \mu +4}{4 \mu ^2 (\mu +1)^3}
				+g_3^2 \frac{2 (\mu +2)}{\mu  (\mu +1)^2}\\
				&-g_1 g_2 \frac{\mu ^3-14 \mu ^2-19 \mu -4}{4 \mu ^2 (\mu +1)^3}
				-g_1 g_3 \frac{8 \mu ^3+29 \mu ^2+22 \mu +1}{2 \mu  (\mu +1)^3}\\
				&-g_2 g_3 \frac{16 \mu ^4+47 \mu ^3+34 \mu ^2+7 \mu +4}{2 \mu ^2 (\mu +1)^3}
				+\frac{g_3 g_2^3}{u} \frac{4 \mu ^2+12 \mu +5}{4 \mu ^2 (\mu +1)^3}\\
				&+\frac{g_1 g_3 g_2^2}{u} \frac{4 \mu ^2+12 \mu +5}{4 \mu ^2 (\mu +1)^3}
				-\frac{g_3^2 g_2^2}{u} \frac{5 \mu ^2+12 \mu +4}{2 \mu  (\mu +1)^3}\\
				&-\frac{g_1 g_3^2 g_2}{u} \frac{5 \mu ^2+12 \mu +4}{2 \mu  (\mu +1)^3}
				+\frac{g_2^4}{u} \frac{1}{8 \mu  (\mu +1)^3}
				+\frac{g_1 g_2^3}{u} \frac{1}{8 \mu  (\mu +1)^3}\\
				&-\frac{g_3^3 g_2}{u} \frac{1}{(\mu +1)^3}
				-\frac{g_1 g_3^3}{u} \frac{1}{(\mu +1)^3}
			\end{aligned} 
		\label{MS_eq:deltaJ} 
	\end{CEquation}	
Inserting the perturbative contributions \eqref{dm} 
- \eqref{MS_eq:Dgammalon}, \eqref{MS_eq:deltagamma1} 
- \eqref{MS_eq:deltagamma3}
and \eqref{MS_eq:deltaJ} in equations \eqref{MS_eq:recursionU}-\eqref{MS_eq:recursiong3} we get the recursion relations of the Malthusian Toner-Tu theory. All the perturbative contributions can be found in the attached "PerturbativeContributions.nb"  Wolfram Mathematica notebook.  

\section{3 - Renormalization group fixed points}
This section discusses some consistency checks that the recursion relations must pass. The Malthusian Toner-Tu theory, just as the Toner-Tu theory, combines the Landau-Ginzburg $\lambda \phi^4$ theory and the Navier-Stokes theory equations of fluid dynamics. The first one accounts for the ferromagnetic-like interaction, while the latter brings the theory off-equilibrium and accounts for the particle's movement at the microscopic level. Therefore,  when we remove all active elements, the theory must reproduce the behavior of equilibrium isotropic and dipolar ferromagnets. Similarly, when we remove all the $\mathcal O(n)$ $\lambda \phi^4$-like terms, the theory must reproduce the behavior of a stirred fluid, see \cite{foster77}. To be even more precise, we can divide the terms of the equation of motion into three groups:

\begin{CEquation} 
\begin{split}
&\mbox{1 - Landau Ginzburg} \qquad \qquad \qquad \qquad r_0 \vv, \ J_0 \vv^2\vv, \ \Gamma_0 \nabla^2 \vv \\
&\mbox{2 - Navier Stokes} \qquad   \qquad \qquad\qquad \qquad   \gamma_1 (\vv\cdot\nabla) \vv, \ \sigma_1\rho_0/ |g^\prime| \nabla \left(\nabla 
\cdot \vv\right)\\
&\mbox{3 - Galilean invariance breaking } \qquad   \quad \    \gamma_2 (\nabla \cdot\vv) \vv,  \gamma_3 \nabla \vv^2
\end{split}
\end{CEquation} 
The first class represents the equilibrium terms of the theory, while the second and third classes contain mostly off-equilibrium, active, terms. 
We shall see how, by removing some ingredients from the Malthusian Toner-Tu theory, we recover many of the know results in the literature; moreover, we will provide two additional consistency checks of the theory. 

\subsection{Equilibrium}
The proper equilibrium limit of the theory is the simple ferromagnet, which can be obtained by neglecting all the terms related to activity. However, the theory admits two other equilibrium limits, which may not have a neat physical interpretation as the isotropic ferromagnet, but that still provide a consistency check for our computation; hence it is reasonable to analyze them. 
\subsubsection{Simple ferromagnet}
If we aim to study the theory in its  \textit{true} equilibrium limit, we must neglect \textit{all} the terms related to activity. This corresponds to neglecting the pressure force $\nabla \left(\nabla \cdot \vv\right)$, the convective derivative term $(\vv\cdot\nabla)\vv$, and all the non-Galilean invariant terms. In this regime the equation of motion takes the form of an equilibrium Langevin equation, 
\begin{CEquation}
    \de_ t \vv= -\Gamma_0 \frac{\delta \mathcal{F}}{\delta \vv} + \boldsymbol f \qquad \qquad \quad 
    \mathcal{F}=\int d^d x \frac{1}{2} (\nabla \vv)^2 + \frac{r_0}{2}\vv^2 + \frac{u_0}{4}\vv^4 \qquad,
    \label{barabba}
\end{CEquation}
where  the force term can be expressed as the derivative of a potential $\mathcal F$. In this limit  the recursion relations \eqref{MS_eq:recursionR}-\eqref{MS_eq:recursiong3} reduce to 
\begin{CAlign}
	r_b & = b^2 r_0 \left[ 1 + \frac{6 u_0}{r_0(r_0+ \Lambda^2)} \ln b \right] & \mu_b &= 1 \\
	u_b &= b^{\epsilon} u_0 \left( 1 -12\, u_0 \ln b \right) & g_{i,b}&=0
	\label{MS_eq:recursiongEq}
\end{CAlign}
which coincide with the ones of an isotropic $\mathcal O(n)$ model with a four-dimensional order parameter. The fixed point of these recursion relations coincides with the Wilson-Fisher fixed point \cite{Wilson1972};
\begin{CAlign}
    r^*&=-\frac{1}{4} \epsilon \Lambda^2 & \mu^*&=1\\
    u^*&=\frac{1}{12}  \epsilon & g_i^*&=0 \quad ,
\end{CAlign}
At the leading order in $\epsilon = 4-d$, this fixed point gives the following critical exponents, 
\begin{CAlign}
    \nu&=\frac{1}{2}+\frac{\epsilon}{8} & \eta&=0 & z&=2
\end{CAlign}
reproducing correctly the critical behavior of an isotropic ferromagnet, or in the classification of Halperin and Hoenberg the critical behavior of model A of critical dynamics \cite{hohenberg1977theory}. 
\subsubsection{Dipolar ferromagnet, and more}
The second model we study is an equilirbrium ferromagnet, in which transverse and longitudinal fluctuations relax differently. The model is described by the follwing equlibrium Langevin equation, 
\begin{CEquation}
 \de_ t \vv= -\Gamma_0 \frac{\delta \mathcal{F}}{\delta \vv} + \boldsymbol f \qquad \qquad 
    \mathcal{F}=\int d^d\vx \frac{1}{2} (\nabla \vv)^2 + \frac{\mu-1}{2} \left(\boldsymbol{\nabla}\cdot\vv\right)^2  + \frac{r_0}{2}\vv^2 + \frac{u_0}{4}\vv^4 \qquad .
    \label{TheOneRing} 
\end{CEquation}

This theory interpolates between an ordinary ferromagnet, $\mu=1$, and a ferromagnet with dipolar interactions, $\mu=\infty$, where the configurations with $\boldsymbol{\nabla}\cdot\vv\neq0$ are completely suppressed (their Boltzmann weight $e^{-\mathcal{F}}$ vanishes). This analogy is explained by the fact that dipolar interactions suppress all the longitudinal fluctuations of $\vv$. 

This theory \eqref{TheOneRing} can be obtained by setting $\gamma_i = 0$. We stress again that the physical interpretation of this theory is not clear - we set to zero some of the parameters related to activity, namely $\gamma_i$, and let the pressure term $\sigma_i$ take any arbitrary value. In this limit the recursion relations of the Malthusian Toner-Tu model reduce to 
\begin{CAlign}
	r_b & = b^2 r_0 \left[ 1 + 3 u_0 \frac{(3 \mu_0+1)\Lambda^2 +4 r_0}{2 r_0 (r_0+\Lambda^2) (r_0+\mu_0\Lambda^2)} \ln b \right] \\
	u_b &= b^{\epsilon} u_0 \left[ 1 - u_0 \frac{17 \mu_0 ^2+2 \mu_0 +5}{2 \mu_0 ^2} \ln b \right]\\
	\mu_b&=\mu_0\\
	g_{i,b}&=0
	\label{MS_eq:recursiongEq2}
\end{CAlign}
If these recursion relations were exact, we would find a manifold of fixed points, namely infinitely many fixed points,  one for each $\mu_0$. However, it is plausible that this is an artifact of perturbation theory at first order in $\epsilon$. Since neither $\Gamma^\perp$ nor $\Gamma^\parallel$ take perturbative corrections at this order, $\mu$ is kept fixed to its bare value along all the RG flow.
This may not be the case when higher-order corrections are taken into account, and thus we shall not discuss further this general scenario. We will only treat the limit $\mu\to\infty$, since in this regime the model should correspond to a dipolar ferromagnet  \cite{aharony73dipolar}, and it does.
The recursion relations for $u$ and $r$ as $\mu\to\infty$ are given by
\begin{CAlign}
	r_b & = b^2 r_0 \left[ 1 +  \frac{9 u_0}{2 r_0 (r_0+\Lambda^2)} \ln b \right] \\
	u_b &= b^{\epsilon} u_0 \left[ 1 - \frac{17}{2} u_0 \ln b \right]
	\label{MS_eq:recursiongEqSol}
\end{CAlign}

The fixed point of these recursion is given, at linear order in $\epsilon$, by
\begin{CAlign}
    r^*&=-\frac{9}{34}\epsilon \Lambda^2 &  u^*&=\frac{2}{17}\epsilon \qquad, 
\end{CAlign}
and the associated critical exponents, in agreement with the literature \cite{aharony73dipolar}, are
\begin{CAlign}
    \nu&=\frac{1}{2}+\frac{9}{68}\epsilon & \eta&=0 & z&=2 \qquad. 
\end{CAlign}
\subsubsection{Another equilibrium model}
The third check, proposed by J. Toner and A. Maitra, consist of considering the following equilibrium theory, 
\begin{CEquation}
  \de_ t \vv= -\Gamma_0 \frac{\delta \mathcal{F}}{\delta \vv} + \boldsymbol f \qquad \qquad 
    \mathcal{F}=\int d^d\vx \frac{1}{2} (\nabla \cdot \vv)^2+ \frac{\mu-1}{2} \left(\boldsymbol{\nabla}\cdot\vv\right)^2 + \frac{r_0}{2}\vv^2 + \frac{u_0}{4}\vv^4 + \lambda_0 \vv\cdot\vv \left(\boldsymbol{\nabla}\cdot\vv\right) \quad , 
    \label{MyPrescious} 
\end{CEquation}
which is a $\lambda \phi^4$ like theory, as the one define in \eqref{TheOneRing}, at which we added the term $\vv^2 (\nabla \cdot \vv)$ in the free energy. We can obtain this theory from the equation of motion of the Malthusian Toner-Tu theory by setting $\gamma_1$ to zero and $\gamma_2 = -2\gamma_3= -2\lambda_0 /\Gamma_0$. Clearly, the fact that this model is at equilibrium depends crucially on the condition $\gamma_2 = - 2 \gamma_3$. This constraint must be preserved by the renormalization group - it would be strange that the large-scale properties of an equilibrium theory revealed to be off-equilibrium.  Requiring that  the renormalization group transformation preserves this constraint, $\gamma_2 = -2\gamma_3$,  implies that the perturbative contributions $\delta \gamma_2$ and $\delta \gamma_3$, defined in equations \eqref{MS_eq:deltagamma2} and \eqref{MS_eq:deltagamma3}, must coincide when  we set $\gamma_1$ to zero and $\gamma_2 = -2 \gamma_3$. We can easily check that this is the case
\begin{CEquation}
    \delta\gamma_2=\delta\gamma_3=g_3^2\frac{3 \mu +4}{\mu ^2}-u \frac{9 \mu ^2-2 \mu +5 }{2 \mu ^2}
\end{CEquation}

\subsection{Off-equilibrium incompressible} 
Here we study the incompressible regime, $\mu \to \infty$ of the Malthusian Toner-Tu theory. In this regime, the theory must reproduce the behavior of incompressible stirred fluids \cite{foster77}, and of incompressible active matter \cite{chen2015critical}.

We can easily convince ourselves that  $\mu = \infty$ corresponds to the incompressible regime. First, 
we recall that $\mu$ is the ratio between longitudinal and transverse kinetic coefficients,
 \begin{CEquation}
 \mu = \frac{\Gamma_0^\parallel}{\Gamma_0^\perp} = 1 +  \sigma_1 \frac{\rho_0}{|g^\prime ( \rho_0) |\Gamma_0 }  \quad ,
 \end{CEquation}
 and, if $\mu = \infty$, longitudinal velocity fluctuations, along with the density fluctuations, are completely suppressed. Furthermore, the difference between longitudinal and transverse kinetic coefficients arises from the pressure force $P =\simeq \sigma_1 \delta \rho$, where $\sigma_1$ is proportional to the inverse compressibility $\chi$: 
\begin{CEquation}
    \chi = \left.\frac{1}{\rho}\frac{\partial \rho}{\partial P}\right|_{\rho_0}=\left. \frac{1}{\rho_0}\frac{\partial \delta\rho}{\partial P}\right|_{\delta\rho=0}=\frac{1}{\rho_0\sigma_1}
\end{CEquation}
Since $\rho_0$ is finite, a system is incompressible if $\sigma_1=\infty$, which corresponds to  $\mu = \infty$. 

We anticipate that any incompressible off-equilibrium fixed point must be infrared-unstable. When $\mu \gg1$, the recursion relation of $\mu$ takes the following form,

\begin{CEquation}
    \mu_b=\mu_0 \left[1+\frac{1}{4} g_{1}^2\ln b + \mathcal O \left ( \frac 1 {\mu_0}\right )\right]\simeq \mu_0 b^{- g_1 ^2 /4} \qquad .
\end{CEquation}
This shows that $\mu=\infty$ is a fixed point; however, since $g_1^2>0$, if $\mu$ is finite, it will flow to smaller and smaller values and escape the incompressible fixed point.
\subsubsection{Incompressible active matter}
The critical behavior of incompressible active matter has been studied by Chen \textit{et al.} in \cite{chen2015critical}.
Here, we limit ourselves to show how to recover the results found in incompressible active matter from the Malthusian Toner-Tu theory. 
The recursion relations of the Malthusian Toner-Tu theory in the incompressible regime,  $\mu= \infty$,  become,

\begin{CAlign} 
	&u_b = b^{\epsilon} u_0 \left[ 1 -\left ( \frac 1 2 g_{1,0}^2 + \frac {17} 2 u_0
	\right) \ln b \right ] \label{pluto} \\
	&r_b=b^2 r_0 \left[1-\left(\frac{9}{2} \frac{r_0-\Lambda^2}{r_0} u_0+\frac{1}{4} g_{1,0}^2\right) \ln b\right]\\
	& g_{1,b} =  b^{\epsilon/2}g_{1,0} \left [ 1-  \left ( \frac 3 8 g_{1,0}^2+\frac 5 3 u_0
	\right) \ln b \right] \label{topolino} \\ 
	& g_{2,b} = b^{\epsilon/2}g_{2,0} \left [ 1 - \left ( \frac 3 8 g_{1,0}^2 + \frac 9 2 u_0 - \frac{11}{12} \frac{g_{1,0} u_0}{g_{2,0}}
	\right) \ln b \right] \\ 
	& g_{3,b} = b^{\epsilon/2}g_{3,0} \left [ 1 - \left (  \frac 3 8 g_{1,0}^2 + \frac 9 2 u_0 - \frac{13}{17} \frac{g_{1,0} u_0}{g_{3,0}}
	\right) \ln b \right]   \label{one} \qquad , 
\end{CAlign}
The recursion relations for $u$, $r$ and $g_1$ \eqref{pluto}-\eqref{topolino} take the same form as in \cite{chen2015critical}, thus leading to the same fixed-point structue. 
Besides the trivial Gaussian fixed point, with $g_i^*=u^*=r^*=0$, and the equilibrium fixed point, which is the dipolar ferromagnet fixed point discussed in the previous section, this incompressible theory has two additional fixed points.

The first fixed point is given by
\begin{CAlign}
    g_1^*&=2 \sqrt{\frac{\epsilon}{3}} &  \mu^*&=\infty\\
    u^*&=0 & r^*&=0  \qquad , 
\end{CAlign}
and it  describes the behavior of incompressible stirred  fluids, namely model B of \cite{foster77}.
Actually, this is not a single fixed point, but a $2$-dimensional manifold of fixed points, since any value of $g_2^*$ and $g_3^*$ represent a distinct fixed point.

The second fixed-point,
\begin{CAlign}
    g_1^*&=2 \sqrt{\frac{31}{113}\epsilon} &  g_{2}^*&=\frac{11}{17}\sqrt{\frac{31}{113}\epsilon} & g_{3}^*&=\frac{13}{17}\sqrt{\frac{31}{113}\epsilon} \\
    \mu^*&=\infty & u^*&=\frac{6}{113}\epsilon & r^*&=-\frac{27}{226}\epsilon\Lambda^2 \quad, 
\end{CAlign}
corresponds to incompressible active matter, at which both advection $g_i$ and ferromagnetic $u$ effective couplings have a finite fixed-point value. The critical exponents of this fixed point coincide with the ones found in \cite{chen2015critical}. We give here the expression of the critical exponents $\nu$ and $z$
\begin{CAlign}
    \nu&=\frac{1}{2}+\frac{29}{226}\epsilon &  z&=2-\frac{31}{113}\epsilon
\end{CAlign}

Interestingly, we managed to derive the critical behavior of incompressible active matter without imposing any constraint on the order parameter. In \cite{chen2015critical} the authors impose a constraint on the order parameter at the beginning of the computation; that is the reason why the equilibrium limit of their model is a dipolar ferromagnet and not a simple isotropic ferromagnet. To some extent, the incompressible limit of the Malthusian Toner-Tu theory looks quite different from the strictly incompressible Toner-Tu Theory, since here we did not impose any constraint.  We remark that it is not trivial that these two theories reproduce the same critical behavior.

\subsection{Galilean invariance}
The last consistency check we shall investigate is given by the Galilean-invariant limit. Galilean invariance is the symmetry under change of reference frame, and it is is defined by the following transformation \cite{landau_fluids}: 
\begin{CEquation}
	\vx^\prime = \vx - \bar \vv t \qquad t^\prime = t \qquad \vv^\prime = \vv + \bar \vv 
	\label{kissmypiss} 
\end{CEquation}
This transformation also implies that the derivatives in the new reference frame are different,
\begin{CEquation}
	\nabla ^\prime = \nabla \qquad \quad \de_{t^\prime} = \de_t - \bar \vv \cdot \nabla  \quad , 
\end{CEquation}
where the prime indicates the derivatives in the new reference frame. 
In general, the TT theory breaks Galilean invariance - in the sense that the transformation \eqref{kissmypiss} is not a symmetry of the equation of motion -  and the Malthusian TT theory is no exception. The reason is that a privileged frame of reference exists, namely that of the surrounding medium in which individuals move. From a more technical point of view, the force terms proportional to $m$, $J$, $\gamma_2$, and $\gamma_3$ explicitly break this symmetry; this means that if we apply the transformation \eqref{kissmypiss} to the equation of motion, we get a theory different from the original one. If we set all these terms to zero, the theory recovers the Galilean symmetry.
 The presence of Galilean invariance, which must be preserved under the RG flow, guarantees that the material derivative $\mathcal{D}_t=\partial_t + (\vv\cdot\boldsymbol{\nabla})$, which is a Galilean-invariant operator, does not take perturbative RG contributions.
As expected, when $m=J=\gamma_2=\gamma_3=0$, or equivalently when $r=u=g_2=g_3=0$, we have that 
 \begin{CEquation} 
    \eta=\delta\gamma_1=0
\end{CEquation}
\subsection{Stability of the RG fixed points}
The stability of a fixed-point is determined by the linear expansion of recursion relations in the neighborhood of the fixed-point.  Following the literature, it is convenient to write the recursion relations in terms of the $\beta$-functions \cite{kardar2007statistical}, which are defined as the derivative of the recursive relations with respect to $\ln b$, at $b=1$.
The $\beta$-function of a generic parameter $\mathcal{P}$ is thus given by,
\begin{CEquation}
    \beta_{\mathcal{P}}=\left.\frac{\partial \mathcal{P}_b}{\partial \ln b}\right|_{b=1}\qquad , 
\end{CEquation}
and its RG recursion relation takes the simple form 
\begin{CEquation}
    \Dot{\mathcal{P}}=\beta_\mathcal{P} \quad , 
    \label{dynsys}
\end{CEquation}
which become equivalent to the recursive relations in the limit of thin shell $\ln b\to0$.

The beta functions for the Malthusian Toner and Tu theory are given by
\begin{CAlign}
	\beta_r & = r \left( 2 + \delta m - \delta \Gamma^\perp \right) \label{MS_eq:betaR}\\
	\beta_\mu &= \mu \left(\delta \Gamma^\parallel  - \delta \Gamma^\perp \right) \label{MS_eq:betaU} \\
	\beta_u &= u \left(\epsilon+\delta J + \delta D -2 \delta \Gamma^\perp -\eta\right)\\
	\beta_{g_{1}} &  = g_{1} \left(\frac{\epsilon}{2}+\delta \gamma_1 + \frac 12 \delta D- \frac 32 \delta \Gamma^\perp-\eta\right)\\
	\beta_{g_{2}} &  = g_{2} \left (\frac{\epsilon}{2}+\delta \gamma_2+ \frac 12 \delta D- \frac 32 \delta \Gamma^\perp-\eta\right)\\
	\beta_{g_{3}} &  = g_{3} \left (\frac{\epsilon}{2}+\delta \gamma_3 + \frac 12 \delta D- \frac 32 \delta \Gamma^\perp-\eta\right)\label{MS_eq:betag3}
\end{CAlign}
where the pertubrative corrections are given by Eq. \eqref{dm} - \eqref{MS_eq:Dgammalon}, \eqref{MS_eq:deltagamma1} - \eqref{MS_eq:deltagamma3} and \eqref{MS_eq:deltaJ}.
In this continuous picture, the fixed-points are determined by the zeros of the $\beta$-functions. It is easy to verify that, at first order in $\epsilon$, all the fixed points we found in the previous sections are zeros of these $\beta$-functions.

The stability of a fixed-point depends on the sign of the eigenvalues of the matrix $H_{ij}$, 
\begin{CEquation} 
H_{ij} = 
\begin{pmatrix} 
\frac{d \beta_r}{d r}  & \frac{d \beta_\mu}{d r}  & \frac{d \beta_u}{d r}  & \frac{d \beta_{g1}}{d r}& \frac{d \beta_{g2}}{d r}& \frac{d \beta_{g3}}{d r}  \\ \\

\frac{d \beta_r}{d \mu}  & \frac{d \beta_\mu}{d \mu}  & \frac{d \beta_u}{d\mu}  & \frac{d \beta_{g1}}{d \mu}& \frac{d \beta_{g2}}{d \mu}& \frac{d \beta_{g3}}{d \mu}  \\ \\

\frac{d \beta_r}{d u}  & \frac{d \beta_\mu}{d u}  & \frac{d \beta_u}{d u}  & \frac{d \beta_{g1}}{du }& \frac{d \beta_{g2}}{d u}& \frac{d \beta_{g3}}{d u}  \\ \\

\frac{d \beta_r}{d g_1}  & \frac{d \beta_\mu}{d g_1}  & \frac{d \beta_u}{d g_1}  & \frac{d \beta_{g1}}{d g_1}& \frac{d \beta_{g2}}{d g_1}& \frac{d \beta_{g3}}{d g_1}  \\ \\

\frac{d \beta_r}{d g_2}  & \frac{d \beta_\mu}{d g_2}  & \frac{d \beta_u}{d g_2}  & \frac{d \beta_{g1}}{d g_2}& \frac{d \beta_{g2}}{d g_2}& \frac{d \beta_{g3}}{d g_2}  \\ \\

\frac{d \beta_r}{d g_3}  & \frac{d \beta_\mu}{d g_3}  & \frac{d \beta_u}{d g_3}  & \frac{d \beta_{g1}}{d g_3}& \frac{d \beta_{g2}}{d g_3}& \frac{d \beta_{g3}}{d g_3}  \\
\end{pmatrix}
\end{CEquation} 
computed at that given fixed-point.  If all eigenvalues are negative the fixed point is stable, and the presence of at least one direction of instability makes the fixed-point unstable.
We must however remind that there is always a direction of instability; this is the direction orthogonal to the critical manifold, namely the direction of the mass $r$. The corresponding eigenvalue determines the runaway exponent, and it is equal to the inverse of the exponent $\nu$.
\subsubsection{Simple ferromagnet}
The simple ferromagnet fixed point is 
\begin{CAlign}
    r^*&=-\frac{\epsilon}{4}\Lambda^2 & u^*&=\frac{\epsilon}{12} & \mu^*&=0 & g_i^*&=0 \quad , 
\end{CAlign}
and the eigenvalues of $H_{ij}$ at this fixed point are, 
\begin{CAlign}
    &2-\frac{\epsilon }{2} & &\frac{\sqrt{151}+5}{48} \epsilon& &-\epsilon & &-\frac{\sqrt{151}-5}{48}  \epsilon & &0 & &0
\end{CAlign}
The eigenvalue $2-\frac{\epsilon}{2}$ is the exponent associated with the parameter $r$, and, as it must be, it is positive. However, there is at least one other direction of instability represented by the eigenvalue $\left(\sqrt{151}+5\right) \epsilon/ 48$, which corresponds to perturbations in the off-equilibrium direction. The presence of two marginally stable directions, with null eigenvalues, may be an artifact of perturbation theory. One of these two directions corresponds to perturbations of $\mu$, which is marginally stable because in the equilibrium limit $\mu$ does not take perturbative at the leading order in $\epsilon$. At the first order in $\epsilon$, we cannot say much about the stability of  $\mu = 0$, since it depends on the higher-order terms in the $\epsilon$-expansion. The second null eigenvalue corresponds to a perturbation of $g_2$ and $g_3$, of the kind $g_2 = -2 g_3 = \lambda$; as we discussed previously this \textit{is not} an off-equilibrium perturbation, because when $g_2 = -2 g_3$ the model is still at equilibrium.  A numerical study of the RG flow in this marginal direction reveals an interesting phenomenology. If we perturb the system along this direction, the RG flow drives $\lambda$ to zero. However, in the meanwhile, that $\lambda \neq 0$ the value of $\mu$ changes, and the point at which the RG flow converges depends on the initial condition. This fixed-point, and the associated critical exponents, interpolate between the simple ferromagnet and the dipolar ferromagnet. Since the fixed point depends on the initial value of the parameters, we should conclude that there is no universality. However, we stress that this is most probably an artifact of the first order expansion in $\epsilon$.

\subsubsection{Dipolar ferromagnet}
The dipolar ferromagnet fixed-point is 
\begin{CAlign}
    r^*&=-\frac{9}{34}\epsilon\Lambda^2 & u^*&=\frac{2}{17}\epsilon & \mu^*&=\infty & g_i^*&=0 \qquad, 
\end{CAlign}
and the eigenvalues of $H_{ij}$ at this fixed-point are
\begin{CAlign}
    &2-\frac{9}{17} \epsilon  & &\frac{31 }{102}\epsilon  & &-\epsilon & &-\frac{\epsilon }{34} & &-\frac{\epsilon }{34} & &0
\end{CAlign}
The eigenvalue $2-\frac{9 \epsilon }{17}$ represents the run-away from the critical surface associated with the relevant parameter $r$.  As the simple ferromagnet, also the dipolar ferromagnet fixed-point is unstable with respect to off-equilibrium perturbations, corresponding to the eigenvalue $\frac{31}{102}\epsilon$. In this case there is only one marginally stable direction, deriving again from the fact that $\mu$ does not take RG corrections, as for the simple ferromagnet fixed point.
The direction $g_2 = -2 g_3 = \lambda$, which was a direction of marginal stability for the simple ferromagnet, is here  a stable  direction.

\subsubsection{Incompressible stirred fluid}
The fixed point for randomly stirred incompressible fluids, given by
\begin{CAlign}
    g_1^*&=2 \sqrt{\frac{\epsilon}{3}} &  \mu^*&=\infty & u^*&=0 & r^*&=0 
\end{CAlign}
results also to be unstable since the eigenvalues of $H_{ij}$ are
\begin{CAlign}
    &2-\frac{\epsilon }{3}  & &\frac{\epsilon }{3} & &\frac{\epsilon }{3} & &-\epsilon & &0 & &0
\end{CAlign}
The exponent $2-\frac{\epsilon }{3}$ rules the run-away in the direction of $r$, while the two eigenvalues $\frac{\epsilon }{3}$ define two additional directions of instability. The former represents an instability with respect to violations of incompressibility - it tells us that for any large but finite value $\mu<\infty$, the RG-flow runs away from this fixed point. The second instability is instead an instability with respect to the presence of any small force $-\left(m_0+J_0 v^2\right)\vv$.
The two marginal directions correspond respectivly to perturbations of $g_2$ and $g_3$. 

\subsubsection{Incompressible active matter}
The incompressible active matter fixed point is 
\begin{CAlign}
    g_1^*&=2 \sqrt{\frac{31}{113}\epsilon} &  g_{2}^*&=\frac{11}{17}\sqrt{\frac{31}{113}\epsilon} & g_{3}^*&=\frac{13}{17}\sqrt{\frac{31}{113}\epsilon} \\
    \mu^*&=\infty & u^*&=\frac{6}{113}\epsilon & r^*&=-\frac{27}{226}\epsilon\Lambda^2 \quad ,
\end{CAlign}
and the eigenvalues of $H_{ij}$ at this fixed-point are
\begin{CAlign}
   &2-\frac{58}{113}\epsilon & &\frac{31}{113} \epsilon & &-\epsilon  & &-\frac{31}{113} \epsilon & &-\frac{17}{113} \epsilon & &-\frac{17}{113} \epsilon
\end{CAlign}
As usual, the first eigenvalue corresponds to the instability of $r$. In addition to this direction of instability, we also have a second unstable direction associated to the eigenvalue $31 \epsilon/113$. This unstable direction involves $\mu$; this means that while this incompressible active matter fixed point is stable if $\mu$ is strictly infinity, it is unstable for any finite $\mu$.
This gives rise to the crossover described in the main text, with a crossover exponent given precisely by the positive eigenvalue $31 \epsilon/113$.

	\bibliographystyle{apsrev4-2}
	\bibliography{Bibbi}